\documentclass[journal]{IEEEtran}

\usepackage{amsmath,amssymb}
\usepackage{amsfonts}
\usepackage{amsbsy}
\usepackage{graphicx}

\usepackage{subfigure}
\usepackage{epstopdf}
\usepackage{lipsum}
\usepackage{multirow}
\usepackage{blindtext}
\usepackage{balance}
\usepackage{xcolor}
\usepackage{amsmath}
\usepackage[linesnumbered,ruled]{algorithm2e}

\begin{document}

\title{Performance Analysis and Optimization of Bidirectional Overlay Cognitive Radio Networks with Hybrid-SWIPT}
\author{{Addanki~Prathima, Devendra~S.~Gurjar,~\IEEEmembership{Member,~IEEE}, Ha~H.~Nguyen,~\IEEEmembership{Senior Member,~IEEE}, and Ajay~Bhardwaj,~\IEEEmembership{Member,~IEEE}}
\thanks{Copyright (c) 2015 IEEE. Personal use of this material is permitted. However, permission to use this material for any other purposes must be obtained from the IEEE by sending a request to pubs-permissions@ieee.org.}
\thanks{*Corresponding author: Devendra S. Gurjar (devendra.gurjar@ieee.org).}
\thanks{Addanki Prathima and Devendra Singh Gurjar are with the Department of Electronics and Communication Engineering, National Institute of Technology, Silchar, Cachar, Assam, 788010, India (e-mails: pratima.addanki@gmail.com, devendra.gurjar@ieee.org)}
\thanks{Ha H. Nguyen is with the Department of Electrical and Computer Engineering, University of Saskatchewan, Saskatoon,  SK S7N 5C5, Canada (e-mail: ha.nguyen@usask.ca).}
\thanks{Ajay Bhardwaj is with School of Computing and Electrical Engineering, Indian Institute of Technology Mandi, 175005, India (e-mail: ajaybhardwaj@ieee.org).}
\thanks{A preliminary version of some part of this paper has been presented at the  IEEE VTC 2018-Fall, Chicago, USA, August 27-30, 2018.}}

%

\maketitle

\begin{abstract}
This paper considers a cooperative cognitive radio network with two primary users (PUs) and two secondary users (SUs) that enables two-way communications of primary and secondary systems in conjunction with non-linear energy harvesting based simultaneous wireless information and power transfer (SWIPT). With the considered network, SUs are able to realize their communications over the licensed spectrum while extending relay assistance to the PUs. The overall bidirectional end-to-end transmission takes place in four phases, which include both energy harvesting (EH) and information transfer. A non-linear energy harvester with a hybrid SWIPT scheme is adopted in which both power-splitting and time-switching EH techniques are used. The SUs aid in relay cooperation by performing an amplify-and-forward operation, whereas selection combining technique is adopted at the PUs to extract the intended signal from multiple received signals broadcasted by the SUs. Accurate outage probability expressions for the primary and secondary links are derived under the Nakagami-$m$ fading environment. Further, the system behavior is analyzed with respect to achievable system throughput and energy efficiency. Since the performance of the considered system is strongly affected by the spectrum sharing factor and hybrid SWIPT parameters, particle swarm optimization is implemented to optimize the system parameters so as to maximize the system throughput and energy efficiency. Simulation results are provided to corroborate the performance analysis and give useful insights into the system behavior concerning various system/channel parameters.
\end{abstract}


\section{Introduction}
\IEEEPARstart{S}{pectrum} scarcity is one of the key challenges in current wireless communications, therefore, improving spectrum efficiency has become a crucial design objective in fifth-generation (5G) wireless communication networks \cite{Saxena}. The applicability of spectrum sharing techniques has been identified for large-scale wireless networks to improve the spectrum utilization and to accommodate a large number of devices \cite{Khan2017}. In particular, cognitive radio networks (CRNs) can resolve the problem of spectrum scarcity by adopting different spectrum sharing techniques such as interweave, overlay, and underlay \cite{Goldsmith2009}. In underlay spectrum sharing, the secondary users (SUs) can utilize the licensed spectrum if the interference caused by their transmissions to the primary system remains below a certain threshold. Moreover, in interweave spectrum sharing, the SUs opportunistically harness free spectrum (spectrum holes) for conducting their communications without causing any interference to the primary communications. Whereas, in overlay spectrum sharing, both PUs and SUs can avail the same licensed band for the transmission of their signals provided the SUs facilitate relay cooperation to the PUs. These techniques help autonomous users to make use of the licensed spectrum for carrying out their communications without affecting the performance of legitimate users.

Another critical design aspect of next-generation wireless communication systems is the network lifetime. As most systems being energy-constraint, there is a need to prolong the network lifetime and making them energy self-sufficient \cite{Sudevalayam}. In many cases, for battery-operated systems, recharging or restoring batteries can become inconvenient and undesirable. Also, conventional energy harvesting (EH) sources, e.g., solar, thermoelectric, etc., are highly intermittent as they rely on the surroundings and environmental conditions \cite{Nasir}. Therefore, a promising solution takes advantage of the fact that energy can be effectively harvested from ambient radio-frequency (RF) signals \cite{Varshney}. As a consequence, simultaneous wireless information and power transfer (SWIPT) is gaining strong interests in both research and industry and is considered as the future of self-sustainable wireless networks.

In RF-based EH, the antenna receives the transmitted signal (RF radiation) and harvests energy using appropriate circuitry \cite{Zhou}, which can then be converted into direct current and stored in a battery. There are two popular approaches, namely power splitting (PS) and time switching (TS), for realizing the SWIPT technology in wireless networks \cite{LWang2017}. In a PS approach, the total power of the received RF signal is divided for EH and information processing (IP), whereas in a TS scheme, the receiving node switches in time for enabling EH and IP operations. It is pointed out, however, that the conventional linear EH model is not practically feasible since an EH circuit comprises of non-linear elements such as diodes, inductors, capacitors etc. Therefore, non-linear energy harvester is a more realistic approach towards energy harvesting \cite{ydong}.

\subsection{Prior Works}

In recent years, SWIPT-based cooperative relaying networks have been studied from the perspectives of link reliability and network lifetime  \cite{TLi}-\cite{SourabhTVT}. Specifically, one-way cooperative relay system is considered in \cite{TLi} and \cite{HLee} whereby energy is harvested at the relay node from the RF signals. Focusing on solving the spectral inefficiency of one-way relay networks, an amplify-and-forward (AF) based two-way relaying system has been considered with SWIPT in \cite{Men}-\cite{Hu}. Further, in \cite{Peng} and \cite{Do}, the authors have employed the decode-and-forward (DF) based relaying strategy for two-way SWIPT-enabled relay systems. Recently, the authors in \cite{SourabhTVT} have analyzed the impact of transceiver hardware impairments on the performance of a DF-based two-way relay network that supports RF energy harvesting.

Apart from cooperative relay networks, research works in \cite{Yin2014}-\cite{Kishore2019} have exploited the advantages of SWIPT in CRNs and cellular networks. Specifically, in \cite{Yin2014}, the authors have proposed a SWIPT-based CRN, whereby the secondary node provides relay cooperation or transmits its own signal in separate phases by extracting power from the RF signals. Different from \cite{Yin2014},  the SU can utilize an overlay mode in \cite{Wang2016} to simultaneously transmit both primary and secondary information signals. Herein, the authors have derived the expressions of outage probability (OP) for both systems under the Rayleigh fading environment. Furthermore, the authors in \cite{Im2015} and \cite{Yang2016} have considered an underlay spectrum sharing scheme along with EH and investigated outage performance of the system. As an extension to the system model of \cite{Im2015} and \cite{Yang2016}, the authors in \cite{Kalamkar2017} have assumed multiple primary interferences and quantified the outage and ergodic capacity for the secondary system. The authors in \cite{Verma2017} have utilized energy harvesting in DF-based one-way cooperative CRN (CCRN) and analyzed outage probability and throughput performance for both primary and secondary systems. A similar model has been adopted in  \cite{Yan2017}, where the authors have introduced a dynamic SWIPT protocol with opportunistic relaying. In \cite{Nguyen2018}, the authors have studied OP performance for a cooperative CRN enabled with energy harvesting under the Nakagami-$m$ fading. Recently, the authors in \cite{Mukherjee} have considered bidirectional transmission in a SWIPT-CCRN system under the Rayleigh fading with the DF relaying strategy and analyzed its outage performance. In this work, only PS-based SWIPT has been considered  for energy harvesting.

The authors in \cite{Zhou2018} have studied energy efficiency (EE) optimization for SWIPT by adopting the PS-based scheme in multiple-input multiple-output (MIMO) two-way AF networks. In their work, the objective was to maximize the EE of the network. Further, in \cite{Tang2018}, the authors aimed to optimize the EE in coordinated multi-point SWIPT heterogeneous networks. Focusing on potential applications of spectrum sharing and RF energy harvesting in IoT, the authors in \cite{DSIoT} have introduced a DF-based SWIPT-enabled CCRN by considering a linear model for EH and analyzed system performance in terms of OP, system throughput, and end-to-end transmission delay. Different from the above-mentioned works, the authors in \cite{Ye2019} have considered a CCRN system with an underlay scheme in which transmission of data from a secondary transmitter to a secondary receiver occurs with the help of multiple relays. The authors then analyzed outage performance for various relaying schemes. Further, an opportunistic ambient backscatter transmission model considering opportunistic spectrum sensing has been proposed in \cite{Kishore2019} for an RF-powered CRN, where the authors have formulated an optimization problem to maximize the energy efficiency of the considered network. Very recently, research works in \cite{Deng2019} and \cite{Singh2020} have addressed the security issue of SWIPT-enabled cognitive radio networks by employing physical layer security techniques. Focusing on relay selection, the authors in \cite{Zhang2020} have utilized a neural network to select the best relay in a two-way SWIPT-enabled CRN.  Most of the previous works discussed above on one-way/two-way CRNs have adopted either a linear EH model with Nakagami-$m$ fading channels, or a non-linear EH model with Rayleigh fading channels and considered a single SWIPT receiver design. To the authors' best knowledge, no work has yet investigated performance of a bidirectional PS-TS SWIPT-based CRN with a non-linear EH circuit over  Nakagami-$m$ fading channels.

\subsection{Main Contributions}

Motivated by the above observations, this paper introduces an AF-based\footnote{While AF and DF are two popular relaying techniques for cooperative communication systems, we focus on AF relaying due to its lower complexity as compared to DF relaying.} bidirectional cognitive radio network. Different from \cite{DSIoT}, a hybrid PS-TS based SWIPT is adopted in this paper. Thus, the obtained analysis can be used for both popular EH receiver designs. Furthermore, unlike \cite{DSIoT}, a non-linear energy harvesting model\footnote{We consider a simplified piecewise linear EH model with two linear segments to make the analysis tractable. It is interesting and left as a future work to carry out the analysis for the more general piecewise linear EH model (with $N$ segments) as introduced in \cite{Shi2020, Liu2020} so that the non-linearity of a practical EH circuit can be captured more accurately.} is considered so that more practical insights can be gathered from the performance analysis. The harvested energy is employed at secondary users (SUs) for facilitating relay assistance to the primary users (PUs) and their own information exchange. Moreover, this paper explores cooperative communications to ensure the quality of service (QoS) of the PUs and enhances the overall spectrum efficiency by enabling bidirectional transmissions of primary and secondary users. Another objective of this paper is to exploit soft computing techniques to maximize the achievable system throughput and energy efficiency of the considered system while maintaining the required QoS of the primary system.

The major contributions are summarized below:
\begin{itemize}
 	\item The paper introduces a non-linear EH-based hybrid SWIPT cognitive radio network in which primary and secondary users can realize their bidirectional transmissions.
 	\item With the introduced scheme, accurate expressions of OP are derived for both primary and secondary users under Nakagami-$m$ fading. Subsequently, expressions of the overall system throughput and energy efficiency are obtained for the delay-limited scenario.
 	\item  The critical value of the spectrum sharing factor is determined to obtain the feasible range of the spectrum sharing factor for which the proposed scheme has lower OP than the direct transmission (i.e., without relay cooperation) to maintain the QoS of the primary system.
 	\item An optimization algorithm based on particle swarm optimization (PSO) has been implemented to obtain optimal system design parameters (i.e. spectrum sharing factor, TS factor, and PS factor) such that the overall system throughput and energy efficiency are maximized.
 	\item The paper discusses the effect of various system/channel parameters on the system performance by means of comprehensive numerical and simulation results.
 \end{itemize}

\subsection{Organization}
The remainder of the paper is organized as follows. In Section \ref{Sys}, the system model and the proposed spectrum sharing scheme are described. Section \ref{Performance} evaluates system performance in terms of OP for both PUs and SUs, system throughput and energy efficiency. The optimization framework is implemented in Section \ref{Optimization}. Numerical results are presented in Section \ref{Num}. Section \ref{Conclusion} concludes the paper.

\textit{Notations}: $\textmd{Pr}[\cdot]$, $f_{X}(\cdot)$, and $F_{X}(\cdot)$ denote respectively the probability, probability density function (PDF) and the cumulative distribution function (CDF) of a random variable $X$. The upper incomplete, the lower incomplete, and the complete gamma functions are represented as $\Gamma[\cdot,\cdot]$, $\Upsilon[\cdot,\cdot]$, and $\Gamma[\cdot]$, respectively \cite[eq. (8.350)]{math}. $\mathcal{K}_{v}(\cdot)$ represents $v^\textmd{th}$ order modified Bessel function of second kind \cite[eq. (8.432.1)]{math} and $\mathcal{W}_{u,v}[\cdot]$ denotes Whittaker function \cite[eq. (9.222)]{math}.

\section{System Model and Scheme}\label{Sys}

\begin{figure}[t]
	\centering
	\includegraphics[width=3.5in]{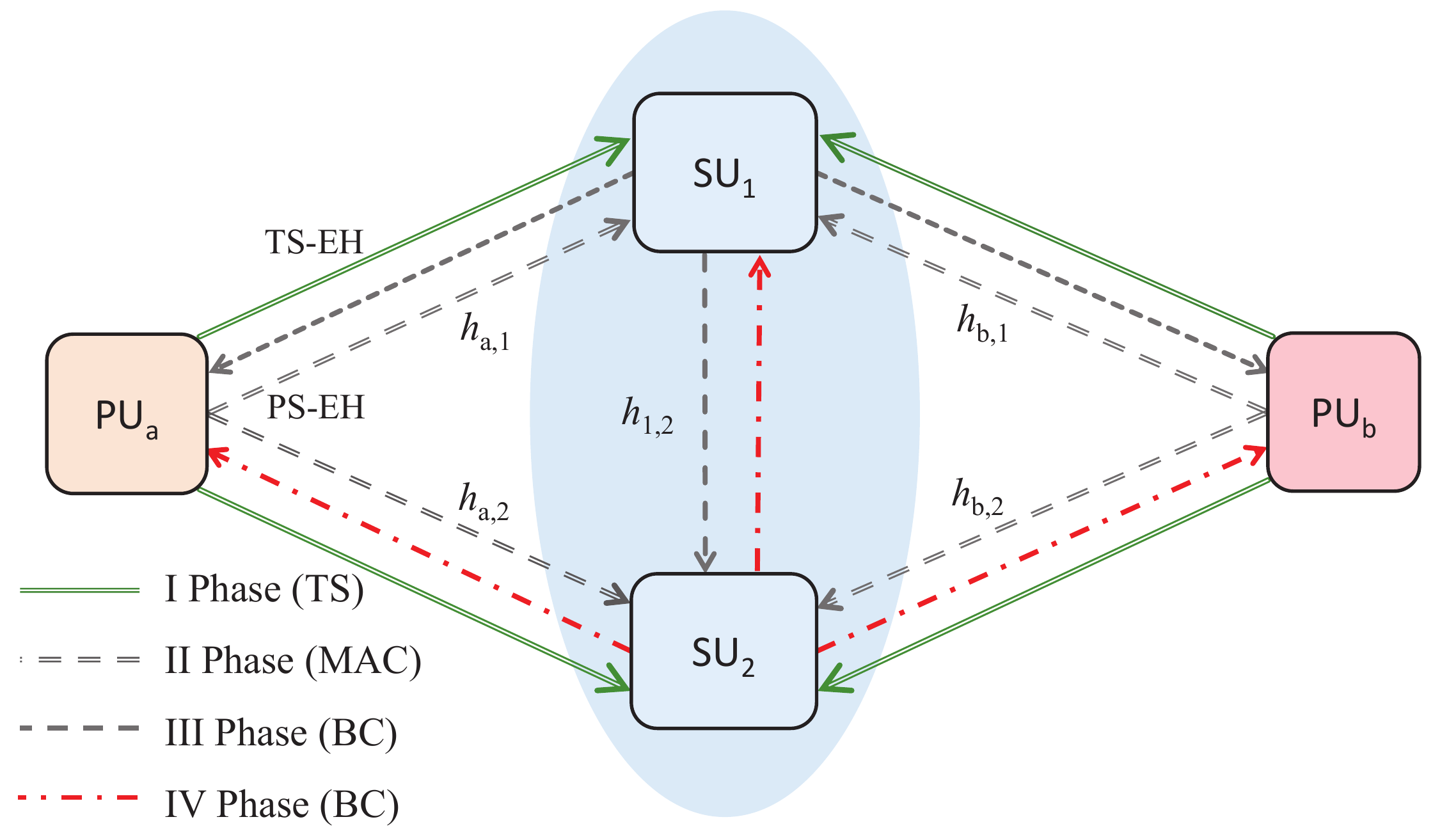}
	\caption{System model for hybrid TS/PS SWIPT in cognitive radio networks.}
	\label{fig01}
\end{figure}

 \begin{figure}[t]
 	\centering
 	\includegraphics[width=3.5in]{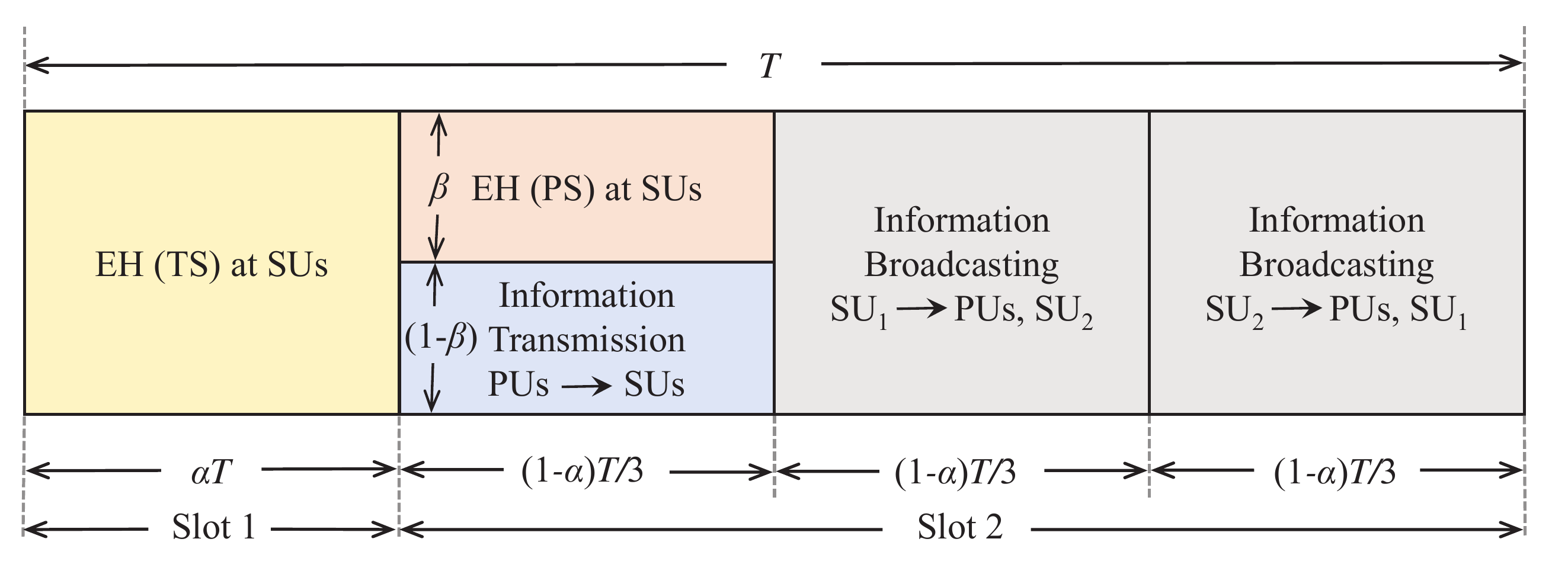}
 	\caption{Transmission block of hybrid-SWIPT in a cooperative CRN.}
 	\label{fig1}
 \end{figure}

As illustrated in Fig. \ref{fig01}, a cooperative CRN enhanced with the SWIPT technology is considered wherein two PUs, denoted as ${\sf PU}_{a}$ and ${\sf PU}_{b}$, wish to exchange information to each other. However, the direct transmission link between the two PUs is not capable enough to provide the required target rates because of heavy shadowing. Therefore, two SUs in their vicinity, denoted as ${\sf SU}_{1}$ and ${\sf SU}_{2}$, provide cooperative relaying of primary users' signals and also carry out their own communications over the same spectrum allocated to the primary system. The PUs and SUs are assumed to be single-antenna devices and operate in a half-duplex mode. The overall EH and bidirectional information exchange between the primary and secondary systems are executed in four phases, as depicted in Fig. \ref{fig1}. Specifically, in the first two phases, the SUs harvest energy using hybrid TS and PS SWIPT. This harvested energy is then stored and used for their mutual information exchange and cooperative relaying to the PUs.

In the first phase, SUs adopt the TS technique to harvest energy from the signals received from the PUs, whereas, in the second phase, they use the PS technique for the same purpose. The remaining fraction of power of the signal received in the second phase is used for information processing and broadcasting. In the third phase, the SU applies the AF technique to the relaying signal and combines it with the signal that is to be transmitted to the other SU. This phase is referred to as broadcast channel (BC) phase. In the fourth phase, which is the second BC phase, the other SU performs the same task as that of the secondary user in the first BC phase (i.e., the third phase). The PUs therefore receive two signal copies from the two SUs, and they apply a selection combination (SC) technique to extract the desired signal.

The system is assumed to operate in a block fading environment wherein the channel coefficients stay constant for the entire duration of the block (comprising of four phases). The channel coefficients are represented as ${h}_{j,i}$ and ${h}_{i,\hat{j}}$, respectively, for the communication links between ${\sf PU}_{j}$ to ${\sf SU}_{i}$ and ${\sf SU}_{i}$ to ${\sf PU}_{\hat{j}}$, for $i\in\{1,2\}$ and $j,\hat{j}\in\{a,b\}$, where $ j\neq\hat{j}$. Likewise, ${h}_{i,\hat{i}}$ denotes the channel gain for the communication links between ${\sf SU}_{i}$ to ${\sf SU}_{\hat{i}}$, where $i\neq\hat{i}$. Further, it is assumed that the channel gains of individual hops are reciprocal, e.g., ${h}_{i,\hat{j}}={h}_{\hat{j},i}$ and ${h}_{i,\hat{i}}={h}_{\hat{i}, i }$. All the signals received at PUs and SUs are affected by additive white Gaussian noise (AWGN). Moreover, the channel coefficient $h_{j,i}$ is considered to follow the Nakagami-$m$ distribution with average power $\Omega_{ij}$ and severity parameter of fading $m_{ij}$.

\subsection{Energy Harvesting}\label{EH}

The structure of one complete block duration of the considered cooperative CRN is shown in Fig. \ref{fig1}. The entire duration $T$ of one block is initially split into two time slots, each of duration $\alpha T$ and $(1-\alpha) T$, respectively, where $0\leq\alpha<1$. The first slot of duration $\alpha T$ is allocated for harvesting energy by the TS approach, whereas the second slot of duration $(1-\alpha) T$ is used for both EH with the PS technique and carrying out the end-to-end bidirectional communications. The energy harvested in the first slot at ${\sf SU}_{i}$ is given as \cite{LiuIET}
\begin{align}\label{ehhh}
\mathcal{E}^{(\textmd{TS})}_{i}=\eta_{i}\alpha T(P_{a}|h_{a,i}|^{2}+P_{b}|h_{b,i}|^{2}),\quad i \in\{1,2\}
\end{align}
where $P_{a}\,\&\,P_{b}$ are the transmit powers at ${\sf PU}_{a}\,\&\,{\sf PU}_{b}$, respectively, and $\eta_{i}$ denotes the efficiency of energy conversion of the EH circuit (in the linear region) at ${\sf SU}_{i}$, where $0<\eta_{i}<1$. The second slot $(1-\alpha) T$ is again split into three phases, each of duration $(1-\alpha) T/3$. In the first slot, both the PUs transmit their signals to the SUs, and is referred to as multiple access channel (MAC) phase. With $x_a$ and $x_b$ being unit-energy symbols transmitted, respectively, by ${\sf PU}_{a}$ and  ${\sf PU}_{b}$ in the MAC phase, the signal received at ${\sf SU}_{i}$ is given as $y_{i}=\sqrt{P_{a}}h_{a,i}x_{a}+\sqrt{P_{b}}h_{b,i}x_{b}+n_{i}$, where $n_{i}\sim\mathcal{CN}(0,\sigma^{2}_{i})$ is the AWGN component at ${\sf SU}_{i}$. The received power at each SU is split into two parts using a power splitting circuit. Specifically, $(\sqrt{\beta}y_{i})^2$ amount of power is utilized for harvesting energy and the remaining $\left(\sqrt{(1-\beta)}y_{i}\right)^2$ power is utilized for completing the information transmission. Thus, the harvested energy at ${\sf SU}_{i}$ in the MAC phase can be expressed as
\begin{align}\label{ehhsh}
\mathcal{E}^{(\textmd{PS})}_{i}=\frac{\beta(1-\alpha) T\eta_{i}}{3}(P_{a}|h_{a,i}|^{2}+P_{b}|h_{b,i}|^{2}).
\end{align}
Considering a non-linear EH model as in \cite{ydong}, from \eqref{ehhh} and \eqref{ehhsh}, the total power transmitted at ${\sf SU}_{i}$ is given as
\begin{align}\label{power}
P_{i}\!\!=\!\!
\begin{cases}
\Delta_{i} (P_{a}|h_{a, i}|^{2}\!+\!P_{b}|h_{b,i}|^{2}), \!\!\!& P_{a}|h_{a,i}|^{2}\!+\!P_{b}|h_{b,i}|^{2} \leq P_{\text{th}} \\
\Delta_{i}P_{\text{th}}, & P_{a}|h_{a,i}|^{2}\!+\!P_{b}|h_{b,i}|^{2}>P_{\text{th}}
\end{cases}
\end{align}
where $\Delta_{i} = \left(\frac{3\eta_{i}\alpha}{1-\alpha}+\beta\eta_{i}\right)$ and $P_{\text{th}}$ is the saturation threshold power of the EH circuit. It is further assumed that the amounts of energy used for signal processing at the relaying SUs are negligible compared to that used for broadcasting the combined signals. Therefore, all the energy harvested at SUs can be utilized for broadcasting operation \cite{Nasir}.
\subsection{Instantaneous Signal-to-Noise Ratios (SNRs)}\label{IP}
The signal received at ${\sf SU}_{i}$ in the MAC phase for information processing and broadcasting, can be written as
\begin{align}\label{sadjlk}
y^{(\textmd{rem})}_{i}=\sqrt{1-\beta}\left(\sqrt{P_{a}}h_{a,i}x_{a}+\sqrt{P_{b}}h_{b,i}x_{b}+n_{i}\right)+n_{c_{i}}
\end{align}
where $n_{c_{i}}\sim\mathcal{CN}(0,\sigma^{2}_{c_{i}})$ represents RF-to-baseband conversion noise. The received signal at ${\sf SU}_{i}$ in MAC phase then undergoes AF operation and is combined with the signal to be transmitted to the other SU. The ${\sf SU}_{i}$ then broadcasts the combined signal. Thus, the broadcast signal in the first BC phase is given as
\begin{align}\label{saadjlk}
x^{(\textmd{BC})}_{i}=\sqrt{\frac{\mu_{i}P_{i}}{\mathcal{G}}}y^{(\textmd{rem})}_{i}+\sqrt{(1-\mu_{i})P_{i}}x_{i}
\end{align}
where $\mu_{i}$ represents the power splitting factor for spectrum sharing and $\mathcal{G}=(1-\beta)(P_{a}|h_{a,i}|^{2}+P_{b}|h_{b,i}|^{2}+\sigma^{2}_{i})$ is the AF gain. Further, the signal received at ${\sf PU}_{j}$, for $j\in\{a,b\}$, can be given as
\begin{align}\label{sadjwalk}
y_{i,j}=\left(\sqrt{\frac{\mu_{i}P_{i}}{\mathcal{G}}}y^{(\textmd{rem})}_{i}+\sqrt{(1-\mu_{i})P_{i}}x_{i}\right)h_{i,j}+n_{j}
\end{align}
where $n_{j}\sim\mathcal{CN}(0,\sigma^{2}_{j})$ represents AWGN variable at ${\sf PU}_{j}$.
The PUs are aware of their transmitted signals and hence they can eliminate the self-interference caused from the received signal to extract the desired signal. Approximating $\mathcal{G}\approx{(1-\beta)(P_{a}|h_{a,i}|^{2}+P_{b}|h_{b,i}|^{2})}$ in \eqref{sadjwalk}, the instantaneous SNR at ${\sf PU}_{j}$ is expressed as
\begin{align}\label{PU_snr}
\gamma_{i,j}=
\begin{cases}
\gamma_{i,j}^{(\text{lin})}, & P_{a}|h_{a, i }|^{2}+P_{b}|h_{b, i }|^{2} \leq P_{\text{th}} \\
\gamma_{i,j}^{(\text{sat})}, & P_{a}|h_{a, i }|^{2}+P_{b}|h_{b, i }|^{2} > P_{\text{th}}
\end{cases}
\end{align}
Further, $\gamma_{i,j}^{(\text{lin})}$ and $\gamma_{i,j}^{(\text{sat})}$ can be expressed as follows.

\begin{align}\label{sdaet}
\gamma_{i,j}^{(\text{lin})}=\frac{\mu_{i}\Delta_{i} P_{\hat{j}}|h_{\hat{j},i}|^{2}}{\omega_{ij}|h_{i,j}|^{2}+\omega_{i\hat{j}}|h_{\hat{j},i}|^{2}+\varepsilon_{ij}}
\end{align}

and

\begin{align}\label{sdaet_sat}
\gamma_{i,j}^{(\text{sat})} = \frac{\mu_{i}\Delta_{i}P_{\text{th}}P_{\hat{j}}|h_{i,j}|^{2}|h_{\hat{j},i}|^{2}}
{\phi_{ij}|h_{i,j}|^{2}+\Phi_{ij}|h_{i,j}|^{4}+\Phi_{i\hat{j}}|h_{i,j}|^{2}|h_{\hat{j},i}|^{2}+
	\varphi_{\hat{j}j}|h_{\hat{j},i}|^{2}}
\end{align}

where $\varepsilon_{ij}=\left(\mu_{i}\Delta_{i}\sigma^{2}_{i}+\frac{\mu_{i}\Delta_{i}\sigma^{2}_{c_{i}}}{1-\beta}\right)$, $\omega_{ij}=(1-\mu_{i})\Delta_{i}P_{j}$, $\omega_{i\hat{j}}=(1-\mu_{i})\Delta_{i}P_{\hat{j}}$, $\phi_{ij}=\left ( \mu_{i} \Delta _{i} P_{\text{th}}\sigma^{2} _{i} + \frac{\mu_{i} \Delta _{i} P_{\text{th}}\sigma^{2} _{c_{i}}}{1-\beta} + P_{j}\sigma^{2} _{j} \right)$, $\Phi_{ij}=\left ( 1-\mu_{i}\right )\Delta _{i} P_{\text{th}}P_{j} $, $\Phi_{i\hat{j}}=\left ( 1-\mu_{i}\right )\Delta _{i} P_{\text{th}}P_{\hat{j}} $ and $\varphi_{\hat{j}j}=P_{\hat{j}}\sigma^{2}_{j}$ for $i\in\{1,2\}$ and $\{j,\hat{j}\}\in\{a,b\}$, with $j\neq\hat{j}$. To obtain the expression of $\gamma^{\textmd{(lin)}}_{i,j}$, the effect of noise terms in the corresponding BC phase is considered negligible as compared to RF-to-baseband conversion noise \cite{HLee} and interference of secondary transmission. Meanwhile, the signal received in the first BC phase at the other SU can be expressed as
\begin{align}\label{sadk}
y_{i,\hat{i}}=\left(\sqrt{\frac{\mu_{i}P_{i}}{\mathcal{G}}}y^{(\textmd{rem})}_{i}+\sqrt{(1-\mu_{i})P_{i}}x_{i}\right)h_{i,\hat{i}}+n_{\hat{i}}
\end{align}
where $n_{\hat{i}}\sim\mathcal{CN}(0,\sigma^{2}_{\hat{i}})$ is the AWGN component at ${\sf SU}_{\hat{i}}$.

The signals transmitted by PUs in the MAC phase are received by both the SUs. Assuming multiuser decoding scenario as adopted in \cite{yli} and \cite{PK2017}, the interference due to primary signals at ${\sf SU}_{\hat{i}}$ in \eqref{sadk} is thus eliminated. Therefore, the instantaneous SNR at ${\sf SU}_{\hat{i}}$ can be expressed as
\begin{align}\label{SU_snr}
\gamma_{i,\hat{i}}=
\begin{cases}
\gamma_{i,\hat{i}}^{(\text{lin})}, & P_{a}|h_{a, i }|^{2}+P_{b}|h_{b, i }|^{2} \leq P_{\text{th}} \\
\gamma_{i,\hat{i}}^{(\text{sat})}, & P_{a}|h_{a, i }|^{2}+P_{b}|h_{b, i }|^{2} > P_{\text{th}}
\end{cases}
\end{align}

where $\gamma_{i,\hat{i}}^{(\text{lin})}$ and $\gamma_{i,\hat{i}}^{(\text{sat})}$ are given as follows.

\begin{align}\label{sadkdd}
\gamma_{i,\hat{i}}^{(\text{lin})}=\frac{\zeta_{i}\left(P_{a}|h_{a, i }|^{2}+P_{b}|h_{b, i }|^{2}\right)|h_{i,\hat{i}}|^{2}}{\xi_{i}|h_{i,\hat{i}}|^{2}+\sigma^{2}_{\hat{i}}}
\end{align}
and

\begin{align}\label{sadkdd_sat}
\gamma_{i,\hat{i}}^{(\text{sat})} = \frac{\Psi_{i}\left(P_{a}|h_{a, i }|^{2}+P_{b}|h_{b, i }|^{2}\right)|h_{i,\hat{i}}|^{2}}{\mathcal{C}_{i}|h_{i,\hat{i}}|^{2}+\mathcal{D}_{\hat{i}}\left(P_{a}|h_{a, i }|^{2}+P_{b}|h_{b, i }|^{2}\right)}
\end{align}

where $\zeta_{i}=(1-\mu_{i})\Delta_{i}$, $\xi_{i}=\mu_{i}\Delta_{i}\left(\sigma^{2}_{i}+\frac{\sigma^{2}_{c_{i}}}{1-\beta}\right)$, $\Psi_{i}=(1-\mu_{i})(1-\beta)\Delta_{i}P_{\text{th}}$, $\mathcal{C}_{i}=\mu_{i}\Delta_{i}P_{\text{th}}\left((1-\beta)\sigma^{2}_{i}+\sigma^{2}_{c_{i}}\right)$ and $\mathcal{D}_{\hat{i}}=(1-\beta)\sigma^{2}_{\hat{i}}$.

Likewise in the second BC phase, ${\sf SU}_{\hat{i}}$ broadcasts its own signal $x_{\hat{i}}$ meant for ${\sf SU}_{i}$ along with the primary signals. The end-to-end instantaneous SNR at each PU for this signal transfer is obtained by replacing $i$ with $\hat{i}$ in \eqref{PU_snr}. Similarly, the instantaneous SNR at ${\sf SU}_{i}$ can be formulated by replacing $i$ with $\hat{i}$ and vice versa in \eqref{SU_snr}.

\section{Performance Analysis}\label{Performance}

\subsection{Outage Probability of the Primary System}\label{OPP}
Outage probability is a crucial performance measure that can assess reliability of communication link of a wireless system operating in a fading channel. In the considered network, the relaying SUs broadcast two signal copies in two BC phases to the PUs. Thus, an outage event at any PU, after considering both the intended signal copies, occurs when the instantaneous data rate is lower than a pre-defined target data rate. Accordingly, the OP for the PUs can be defined as
\begin{align}\label{oopc}
 \mathcal{P}_{\textmd{out},j}&=\textmd{Pr}\left[\mathcal{R}_{\textmd{sc},j}<r_{j}\right]
 =\textmd{Pr}\left[\max\left(\gamma_{i,j},\gamma_{\hat{i},j}\right)<\gamma_{j}\right]
 \end{align}
 where the instantaneous rate $\mathcal{R}_{\text{sc},j}=\frac{1-\alpha}{3}\log_{2}\left(1+\max\left(\gamma_{i,j},\gamma_{\hat{i},j}\right)\right)$ and $\gamma_{j}=2^{\frac{3r_{j}}{1-\alpha}}-1$, for $j\in\{a,b\}$, $i,\hat{i}\in\{1,2\}$ and $i\neq\hat{i}$. Further,  \eqref{oopc} can be expressed as
 \begin{align}\label{sjeo}
 \mathcal{P}_{\textmd{out},j}=\prod^{2}_{i=1}F_{\gamma_{i,j}}(\gamma_{j})
 \end{align}
 where
  \begin{align}\label{Pouta}
 F_{\gamma_{i,j}}(\gamma_{j})=F_{\gamma_{i,j}^{(\text{lin})}}(\gamma_{j})+F_{\gamma_{i,j}^{(\text{sat})}}(\gamma_{j}).
 \end{align}
Accurate expressions of the CDF of $F_{\gamma_{i,j}^{(\text{lin})}}(\gamma_{j})$ and $F_{\gamma_{i,j}^{(\text{sat})}}(\gamma_{j})$ are provided in the following lemma.

 \newtheorem{lemma}{Lemma}
  \begin{lemma}\label{lem1}
 	The expression of $F_{\gamma^{(\text{lin})}_{i,j}}(\gamma_{j})$, for $j\in\{a,b\}$ can be presented as
 	\begin{align}\label{saljd}
 	\!\!\!F_{\gamma^{(\text{lin})}_{i,j}}(\gamma_{j})&\!=\!\frac{\Gamma [m_{ij}]-\Gamma \left[m_{ij},\frac{m_{ij} P_\text{th}}{\Omega_{ij} P_{j} }\right]}{\Gamma [m_{ij}]} - \frac{\left(\frac{m_{ij}}{\Omega_{ij}}\right)^{m_{ij}}}{\Gamma [m_{ij}]}
 	\mathrm{e}^{\left(-\frac{m_{i\hat{j}} P_\text{th}}{\Omega_{i\hat{j}} P_{\hat{j}}} \right)}\nonumber\\
 	&\!\times\!\!
 	\sum _{k=0}^{m_{i\hat{j}-1}} \frac{\left(\frac{m_{i\hat{j}}}{\Omega_{i\hat{j}} P_{\hat{j}}}\right)^k}{k!} \sum _{q=0}^{k} P_\text{th}^{q} \binom{k}{q} (-P_{j})^{k-q}\nonumber\\
 	&\!\times\!\!
 	\sum _{t=0}^{k+m_{ij}-q-1} \frac{(-1)^t t! \binom{k+m_{ij}-q-1}{t}}{\left(\frac{m_{i\hat{j}} P_{j}}{\Omega_{i\hat{j}} P_{\hat{j}}}-\frac{m_{ij}}{\Omega_{ij}}\right)^{t+1}}\nonumber\\
 	&\!\times\!\!
 	\left(\left(\frac{P_\text{th}}{P_{j}}\right)^{k+m_{ij}-q-t-1} \mathrm{e}^{\frac{P_\text{th}}{P_{j}}\left(\frac{m_{i\hat{j}} P_{j}}{\Omega_{i\hat{j}} P_{\hat{j}}}-\frac{m_{ij}}{\Omega_{ij}}\right)}\right.\nonumber\\
 	&- \left.\iota_1^{k+m_{ij}-q-t-1} \mathrm{e}^{\iota_1 \left(\frac{m_{i\hat{j}} P_{j}}{\Omega_{i\hat{j}} P_{\hat{j}}}-\frac{m_{ij}}{\Omega_{ij}}\right)} \right)\nonumber\\
 	&-\frac{\left(\frac{m_{ij}}{\Omega_{ij}}\right)^{m_{ij}}}{\Gamma [m_{ij}]} \mathrm{e}^{\left(-\frac{m_{i\hat{j}} \varepsilon_{ij} \gamma_{j}}{\Omega_{i\hat{j}} \Xi}\right)}\sum _{p=0}^{m_{i\hat{j}}-1} \frac{\left(\frac{m_{i\hat{j}}}{\Omega_{i\hat{j}} \Xi}\right)^p}{p!}\nonumber\\
 	&\!\times\!\!
 	\sum _{n=0}^p \varepsilon_{ij}^n \gamma_{j}^p \binom{p}{n} \omega_{ij}^{p-n}
 	\left(\frac{\omega_{ij} \gamma_{j} m_{i\hat{j}}}{\Omega_{i\hat{j}} \Xi}\!+\!\frac{m_{ij}}{\Omega_{ij}}\right)^{\!-\!(m_{ij}\!-\!n\!+\!p)}\nonumber\\
 	&\!\times\!\!
 	 \Upsilon\left[m_{ij}-n+p,\left(\frac{m_{ij}}{\Omega_{ij}}+\frac{\omega_{ij}
 		\gamma_{j} m_{i\hat{j}}}{\Omega_{i\hat{j}} \Xi}\right) \iota_1 \right]
  	\end{align}
 	
 	where $\iota_1=\frac{P_\text{th} \Xi -\varepsilon_{ij} \gamma_{j} P_{\hat{j}}}{\omega_{ij} \gamma_{j} P_{\hat{j}}+P_{j} \Xi}$ and $\Xi=\mu_{i}\Delta_{i} P_{\hat{j}}-\omega_{i\hat{j}}\gamma_{j}$. On the other hand, the expression for $F_{\gamma^{(\text{sat})}_{i,j}}(\gamma_{j})$ is as follows:
\begin{align}\label{Poutjsat}
\!\!\!F_{\gamma^{(\text{sat})}_{i,j}}(\gamma_{j}) &= \sum _{p=0}^{m_{ij}-1} \frac{\left(\frac{m_{ij}}{\Omega_{ij} P_{j}}\right)^p}{p!}\sum _{n=0}^p \binom{p}{n}P_\text{th}^n (-P_{\hat{j}})^{p-n}\frac{\left(\frac{m_{i\hat{j}}}{\Omega_{i\hat{j}}}\right)^{m_{i\hat{j}}}}{\Gamma [m_{i\hat{j}}]}\nonumber\\
&\!\times\!\!
\mathrm{e}^ {-\frac{m_{ij} P_\text{th}}{\Omega_{ij} P_{j}}}\sum _{s=0}^{\infty} \frac{\left(\frac{m_{ij} P_{\hat{j}}}{\Omega_{ij} P_{j}}\right)^s}{s!}\Upsilon\left[m_{i\hat{j}}-n+p+s,\frac{m_{i\hat{j}} \iota_2}{\Omega_{i\hat{j}}}\right]\nonumber\\
&\!\times\!\!
\left(\frac{m_{i\hat{j}}}{\Omega_{i\hat{j}}}\right)^{-(m_{i\hat{j}}-n+p+s)}\nonumber\\
&+ \mathrm{e}^{\left(\frac{\mathcal{T}_1 m_{ij}}{\mathcal{T}_2 \Omega_{ij}}\right)}\frac{\left(\frac{m_{i\hat{j}}}{\Omega_{i\hat{j}}}\right)^{m_{i\hat{j}}}}{\Gamma [m_{i\hat{j}}]}\sum _{k=0}^{m_{ij}-1}\frac{\left(\frac{m_{ij}}{\mathcal{T}_2 \Omega_{ij}}\right)^k}{k!}\sum _{q=0}^k\binom{k}{q}\nonumber\\
&\!\times\!\!
(-\mathcal{T}_1)^{k-q} \left(\frac{m_{ij}}{\mathcal{T}_2 \Omega_{ij}} + \frac{m_{i\hat{j}}}{\Omega_{i\hat{j}}}\right)^{-(m_{i\hat{j}}+q)}\nonumber\\
&\!\times\!
\Gamma \left[m_{i\hat{j}}+q,\iota_2 \left(\frac{m_{ij}}{\mathcal{T}_2 \Omega_{ij}} + \frac{m_{i\hat{j}}}{\Omega_{i\hat{j}}}\right)\right]
\end{align}
where $\iota_2=\frac{ \mathcal{T}_2 P_\text{th}+ \mathcal{T}_1 P_{j}}{P_{j}+ \mathcal{T}_2 P_{\hat{j}}}$.
 \end{lemma}
 \begin{IEEEproof}
 	Please see Appendix \ref{appA}.
 \end{IEEEproof}

Inserting \eqref{saljd} and \eqref{Poutjsat} into \eqref{sjeo}, we can get the required expression of OP for the primary system.

\subsection{Outage Probability of the Secondary System}\label{OPS}

Outage occurs at the SU if the instantaneous data rate achieved at that node falls below a pre-defined target data rate. Therefore, the OP at the SU is given as
 \begin{align}\label{oopd}
 \mathcal{P}_{\textmd{out},\hat{i}}=\textmd{Pr}[\mathcal{R}_{i,\hat{i}}< r_{\hat{i}}]
 \end{align}
 for $i,\hat{i}\in\{1,2\}$ and $i\neq\hat{i}$, where $r_{\hat{i}}$ is the target rate required at ${\sf SU}_{\hat{i}}$. Moreover, $\mathcal{R}_{i,\hat{i}}=((1-\alpha)/3)\log_{2}(1+\gamma_{i,\hat{i}})$ is the instantaneous rate achieved at ${\sf SU}_{\hat{i}}$.
 $\textmd{Pr}[\mathcal{R}_{i,\hat{i}}< r_{\hat{i}}]$ in \eqref{oopd} can be formulated as,
 \begin{align}\label{sjfsjgl}
 \textmd{Pr}[\mathcal{R}_{i,\hat{i}}< r_{\hat{i}}]=F_{\gamma_{i,\hat{i}}}(\gamma_{\hat{i}})
 \end{align}
 and
 \begin{align}\label{Pout1}
 F_{\gamma_{i,\hat{i}}}(\gamma_{\hat{i}})=F_{\gamma_{i,\hat{i}}^{(\text{lin})}}(\gamma_{\hat{i}})+F_{\gamma_{i,\hat{i}}^{(\text{sat})}}(\gamma_{\hat{i}})
\end{align}
 where $\gamma_{\hat{i}}=2^{\frac{3r_{\hat{i}}}{1-\alpha}}-1$. The required CDF expression $F_{\gamma_{i,\hat{i}}}(\gamma_{\hat{i}})$ is provided in the following Lemma.
 \newtheorem{lemmaa}{Lemma}
 \begin{lemma}\label{lem2}
 	The CDF $F_{\gamma_{i,\hat{i}}^\text{lin}}(\gamma_{\hat{i}})$ can be expressed as
 	{\small{\begin{align}\label{slkskdl} 			
 		 \!\!\!F_{\gamma^{(\text{lin})}_{i,\hat{i}}}(\gamma_{\hat{i}})&\!=\! \lambda \frac{\Upsilon \left[m_{i\hat{i}},\frac{m_{i\hat{i}} \iota_3}{\Omega_{i\hat{i}}}\right]}{\Gamma [m_{i\hat{i}}]}\left( \Lambda\left(k,\left(\frac{m_{ij}}{\Omega_{ij} P_{j}}\right)\right)-\!\!\!\!\!\!\sum _{t=0}^{-k+m_{ij}+m_{i\hat{j}}+s-2} \right.\nonumber\\
 		&\times \left.  \frac{\left(\frac{m_{i\hat{j}}}{\Omega_{i\hat{j}} P_{\hat{j}}}\right)^t}{t!} \Lambda\left(k+t,\left(\frac{m_{ij}}{\Omega_{ij} P_{j}}+\frac{m_{i\hat{j}}}{\Omega_{i\hat{j}} P_{\hat{j}}}\right)\right)\right)\nonumber\\
        & +  \lambda \left(  \rho\left(k,\left(\frac{m_{ij}}{\Omega_{ij} P_{j}}\right)\right)-\sum _{t=0}^{-k+m_{ij}+m_{i\hat{j}}+s-2} \right.\nonumber\\
        &\times \left.  \frac{\left(\frac{m_{i\hat{j}}}{\Omega_{i\hat{j}} P_{\hat{j}}}\right)^t}{t!} \rho\left(k+t,\left(\frac{m_{ij}}{\Omega_{ij} P_{j}}+\frac{m_{i\hat{j}}}{\Omega_{i\hat{j}} P_{\hat{j}}}\right)\right) \right)\nonumber\\
 		\end{align}}}	
 where $\lambda$, $\Lambda\left(k,\left(\frac{m_{ij}}{\Omega_{ij} P_{j}}\right)\right)$, and $\rho\left(k,\left(\frac{m_{ij}}{\Omega_{ij} P_{j}}\right)\right)$ are defined on the next page in \eqref{lambda}, \eqref{Lambda}, and \eqref{rhofunc}, respectively and $\iota_3=\frac{\gamma_{\hat{i}} \sigma_{\hat{i}}^2}{\zeta_{i} P_\text{th}-\xi_{i} \gamma_{\hat{i}}}$.

 \begin{figure*}[t]
 	{
 	{\begin{align}\label{lambda} 				
    \lambda&=\frac{\left(\frac{m_{ij}}{\Omega_{ij} P_{j}}\right)^{m_{ij}} \left(\frac{m_{i\hat{j}}}{\Omega_{i\hat{j}}P_{\hat{j}}}\right)^{m_{i\hat{j}}}}{\Gamma [m_{ij}] \Gamma [m_{i\hat{j}}]}\sum _{k=0}^{m_{ij}-1} (-1)^{-k+m_{ij}-1} \binom{m_{ij}-1}{k} \sum _{s=0}^{\infty} \frac{\left(\frac{m_{ij}}{\Omega_{ij} P_{j}}\right)^s}{s!} \left(\frac{m_{i\hat{j}}}{\Omega_{i\hat{j}} P_{\hat{j}}}\right)^{-(-k+m_{ij}+m_{i\hat{j}}+s-1)}\nonumber\\
    &\times \Gamma[-k+m_{ij}+m_{i\hat{j}}+s-1]
 			\end{align}}}
 	\hrulefill
 \end{figure*}

  \begin{figure*}[t]
 	{
 		{\begin{align}\label{Lambda} 				
 		\Lambda\left(k,\left(\frac{m_{ij}}{\Omega_{ij} P_{j}}\right)\right) = \left(\frac{m_{ij}}{\Omega_{ij} P_{j}}\right)^{-(k+1)} \Upsilon\left[k+1,\left(\frac{m_{ij} }{\Omega_{ij} P_{j}}\right)P_{\text{th}}\right]
 			\end{align}}}
 	\hrulefill
 \end{figure*}

  	\begin{figure*}[t]
 	{
 		{\begin{align}\label{rhofunc} 				
 			\rho\left(k,\left(\frac{m_{ij}}{\Omega_{ij} P_{j}}\right)\right) &\!=\!  \left( \Gamma[k+1] \left(\frac{m_{ij}}{\Omega_{ij} P_{j}}\right)^{-(k+1)} \frac{\Gamma \left[m_{i\hat{i}},\frac{m_{i\hat{i}} \iota_3}{\Omega_{i\hat{i}}}\right]}{\Gamma [m_{i\hat{i}}]} - \Gamma[k+1] \left(\frac{m_{ij}}{\Omega_{ij} P_{j}}\right)^{-(k+1)} \mathrm{e}^{\left(-\frac{\xi_{i} \gamma_{\hat{i}}}{\zeta_{i}}\right) \left(\frac{m_{ij}}{\Omega_{ij} P_{j}}\right)}
 			\sum _{q=0}^{k} \frac{\left(\frac{m_{ij}}{\Omega_{ij} P_{j}}\right)^q}{q!}\right.\nonumber\\
 			&\times \frac{\left(\frac{m_{i\hat{i}}}{\Omega_{i\hat{i}}}\right)^{m_{i\hat{i}}}}{\Gamma [m_{i\hat{i}}]}\frac{1}{\zeta_{i}^q} \sum _{p=0}^{q}\binom{q}{p}(\xi_{i} \gamma_{\hat{i}})^p \left(\gamma_{\hat{i}} \sigma_{\hat{i}}^2\right)^{q-p} \left(2 \left(\frac{m_{i\hat{i}}}{\Omega_{i\hat{i}}}\right)^{-\frac{1}{2} (m_{i\hat{i}}+p-q)} \left(\left(\frac{\gamma_{\hat{i}} \sigma_{\hat{i}}^2}{\zeta_{i}}\right) \left(\frac{m_{ij}}{\Omega_{ij} P_{j}}\right)\right)^{\frac{1}{2} (m_{i\hat{i}}+p-q)}\right.\nonumber\\
&\!\times\!\!
 \mathcal{K}_{m_{i\hat{i}}+p-q}\left(2 \sqrt{\frac{m_{i\hat{i}} \left(\gamma_{\hat{i}} \sigma_{\hat{i}}^2\right) \left(\frac{m_{ij}}{\Omega_{ij} P_{j}}\right)}{\zeta_{i} \Omega_{i\hat{i}}}}\right) - \sum _{u=0}^{\infty}\frac{\left(-\frac{m_{i\hat{i}}}{\Omega_{i\hat{i}}}\right)^u}{u!}\left(\left(\frac{\gamma_{\hat{i}} \sigma_{\hat{i}}^2}{\zeta_{i}}\right) \left(\frac{m_{ij}}{\Omega_{ij} P_{j}}\right)\right)^{(m_{i\hat{i}}+p+u-q)}\nonumber\\
&\!\times\! \left.
 \left(\left(\frac{\gamma_{\hat{i}} \sigma_{\hat{i}}^2}{\zeta_{i}}\right) \left(\frac{m_{ij}}{\Omega_{ij} P_{j}}\right) \frac{1}{\iota_3}\right)^{-\frac{1}{2}(m_{i\hat{i}}+p+u-q+1)} \mathrm{e}^{-\left(\frac{\gamma_{\hat{i}} \sigma_{\hat{i}}^2}{\zeta_{i}}\right) \left(\frac{m_{ij}}{\Omega_{ij} P_{j}}\right) \frac{1}{2\iota_3}} \right. \nonumber\\
&\!\times\!\! \left. \left.
\mathcal{W}_{-\frac{1}{2}(m_{i\hat{i}}+p+u-q+1), \frac{1-(m_{i\hat{i}}+p+u-q+1)}{2}}\left[\left(\frac{\gamma_{\hat{i}} \sigma_{\hat{i}}^2}{\zeta_{i}}\right) \left(\frac{m_{ij}}{\Omega_{ij} P_{j}}\right) \frac{1}{2\iota_3}\right] \right) \right)	\nonumber\\
 			\end{align}}}
 	\hrulefill
 \end{figure*}
The CDF $F_{\gamma_{i,\hat{i}}^{(\text{sat})}}(\gamma_{\hat{i}})$ can be expressed as,
{\small{\begin{align}\label{Su_sat_1} 	
		\!\!\!F_{\gamma_{i,\hat{i}}^{(\text{sat})}} (\gamma_{\hat{i}})&\!=\!\lambda\left(\varrho\left(k,\left(\frac{m_{ij}}{\Omega_{ij} P_{j}}\right)\right)-\sum _{t=0}^{-k+m_{ij}+m_{i\hat{j}}+s-2} \right.\nonumber\\
		&\times \left.  \frac{\left(\frac{m_{i\hat{j}}}{\Omega_{i\hat{j}} P_{\hat{j}}}\right)^t}{t!} \varrho\left(k+t,\left(\frac{m_{ij}}{\Omega_{ij} P_{j}}+\frac{m_{i\hat{j}}}{\Omega_{i\hat{j}} P_{\hat{j}}}\right)\right)\right. \nonumber\\
		&+ \left. \epsilon\left(k,\left(\frac{m_{ij}}{\Omega_{ij} P_{j}}\right)\right)-\sum _{t=0}^{-k+m_{ij}+m_{i\hat{j}}+s-2} \frac{\left(\frac{m_{i\hat{j}}}{\Omega_{i\hat{j}} P_{\hat{j}}}\right)^t}{t!} \right.\nonumber\\
		&\times \left.
		\epsilon\left(k+t,\left(\frac{m_{ij}}{\Omega_{ij} P_{j}}+\frac{m_{i\hat{j}}}{\Omega_{i\hat{j}} P_{\hat{j}}}\right)\right)\right)
		\end{align}}}

where $\iota_4=\frac{\mathcal{D}_{\hat{i}} \gamma_{\hat{i}} P_\text{th}}{\Psi_{i} P_\text{th}-\mathcal{C}_{i} \gamma{\hat{i}}}$. $\varrho\left(k,\left(\frac{m_{ij}}{\Omega_{ij} P_{j}}\right)\right)$ and $\epsilon\left(k,\left(\frac{m_{ij}}{\Omega_{ij} P_{j}}\right)\right)$ are defined in \eqref{I1sat} and \eqref{P2}, respectively.	

 	\begin{IEEEproof}
 		Please see Appendix \ref{appB}.
 	\end{IEEEproof}
 \end{lemma}


		\begin{figure*}[t]
		{
			{\begin{align}\label{I1sat}
			\varrho\left(k,\left(\frac{m_{ij}}{\Omega_{ij} P_{j}}\right)\right) &\!=\!\left(\frac{m_{ij}}{\Omega_{ij} P_{j}}\right)^{-(k+1)} \left( k! \frac{\Upsilon\left[m_{i\hat{i}},\frac{m_{i\hat{i}}}{\Omega_{i\hat{i}}} \iota_4 \right] - \Upsilon \left[m_{i\hat{i}},\frac{m_{i\hat{i}}}{\Omega_{i\hat{i}}} \frac{\mathcal{D}_{\hat{i}} \gamma_{\hat{i}}}{\Psi_{i}}\right]}{\Gamma[m_{i\hat{i}}]} - k! \frac{\left(\frac{m_{i\hat{i}}}{\Omega_{i\hat{i}}}\right)^{m_{i\hat{i}}}}{\Gamma [m_{i\hat{i}}]} \sum _{p=0}^{k} \frac{\left(\frac{m_{ij}}{\Omega_{ij} P_{j}}\right)^p}{p!}(\mathcal{C}_{i} \gamma_{\hat{i}})^p \frac{1}{\Psi_{i}^{m_{i\hat{i}}+p}} \right. \nonumber\\
			& \!\times\! \left.
			\mathrm{e}^{-\left(\left(\frac{m_{ij}}{\Omega_{ij} P_{j}}\right) \frac{\mathcal{C}_{i} \gamma_{\hat{i}}}{\Psi_{i} }+\frac{m_{i\hat{i}}}{\Omega_{i\hat{i}}} \frac{\mathcal{D}_{\hat{i}} \gamma_{\hat{i}}}{\Psi_{i}}\right)} \!\!\sum _{q=0}^{m_{i\hat{i}}+p-1} \binom{m_{i\hat{i}}+p-1}{q}  (\mathcal{D}_{\hat{i}} \gamma_{\hat{i}})^q \sum _{u=0}^{\infty} \frac{\left(-\frac{m_{i\hat{i}}}{\Psi_{i} \Omega_{i\hat{i}}}\right)^u}{u!} \left(\left(\frac{m_{ij}}{\Omega_{ij} P_{j}}\right) \frac{\mathcal{C}_{i} \gamma_{\hat{i}}}{\Psi_{i}} \mathcal{D}_{\hat{i}} \gamma_{\hat{i}} \right)^{m_{i\hat{i}}-q+u} \right. \nonumber\\
			& \times \left.
			\left(\left(\frac{m_{ij}}{\Omega_{ij} P_{j}}\right) \frac{\mathcal{C}_{i} \gamma_{\hat{i}}}{\Psi_{i}} 	 \frac{\mathcal{D}_{\hat{i}} \gamma_{\hat{i}}}{\Psi_{i} \iota_4 -\mathcal{D}_{\hat{i}} \gamma_{\hat{i}}} \right)^{-\frac{1}{2} (m_{i\hat{i}}-q+u+1)} 	 \mathrm{e}^{-\frac{1}{2}\left(\frac{m_{ij}}{\Omega_{ij} P_{j}}\right) \frac{\mathcal{C}_{i} \gamma_{\hat{i}}}{\Psi_{i}}
				\frac{\mathcal{D}_{\hat{i}} \gamma_{\hat{i}}}{\Psi_{i} \iota_4 -\mathcal{D}_{\hat{i}} \gamma_{\hat{i}}}} \right. \nonumber\\
				& \times \left.
				 \mathcal{W}_{-\frac{1}{2} (m_{i\hat{i}}-q+u+1),\frac{1}{2} (1-(m_{i\hat{i}}-q+u+1))}\left[\left(\frac{m_{ij}}{\Omega_{ij} P_{j}}\right) \frac{\mathcal{C}_{i} \gamma_{\hat{i}}}{\Psi_{i}}
				\frac{\mathcal{D}_{\hat{i}} \gamma_{\hat{i}}}{\Psi_{i} \iota_4 -\mathcal{D}_{\hat{i}} \gamma_{\hat{i}}} \right] \right.\nonumber\\
				&- \left. \Upsilon\left[k+1, \left(\frac{m_{ij}}{\Omega_{ij} P_{j}}\right) P_\text{th}\right]  \frac{\Upsilon \left[m_{i\hat{i}},\frac{m_{i\hat{i}} \iota_4}{\Omega_{i\hat{i}}}\right] - \Upsilon \left[m_{i\hat{i}},\frac{m_{i\hat{i}}}{\Omega_{i\hat{i}}} \frac{\mathcal{D}_{\hat{i}} \gamma_{\hat{i}}}{\Psi_{i}}\right]}{\Gamma[m_{i\hat{i}}]} \right)
			\end{align}}}
		\hrulefill
	\end{figure*}

	\begin{figure*}[t]
		{
			{\begin{align}\label{P2}
				\epsilon\left(k,\left(\frac{m_{ij}}{\Omega_{ij} P_{j}}\right)\right) &=\left(\frac{m_{ij}}{\Omega_{ij} P_{j}}\right)^{-(k+1)}		\Gamma \left[k+1,\frac{m_{ij}}{\Omega_{ij} P_{j}} P_\text{th}\right]
				\frac{ \Upsilon\left[m_{i\hat{i}},\frac{m_{i\hat{i}}}{\Omega_{i\hat{i}}} \frac{\mathcal{D}_{\hat{i}} \gamma_{\hat{i}}}{\Psi_{i}}\right] }{\Gamma[m_{i\hat{i}}]}
		\end{align}}}
\hrulefill
\end{figure*}			 	
		

\subsection{Spectrum Sharing}\label{Spectrum}
The secondary system employed for relay cooperation should assist in providing better OP at the primary nodes compared to direct end-to-end transmission. Hence, it is useful to analyze the range of spectrum sharing factor such that the system shows better OP than that of the direct link. When the PUs are communicating directly without the help of relaying SUs, the rate achieved at the PUs can be given as
\begin{align}
\mathcal{R}_{j,\hat{j}}^{(\textmd{D})}=\frac{1}{2}\log_{2}\left ( 1+\frac{P_{j}\left | h_{j,\hat{j}} \right |^{\textmd{2}}}{\sigma ^{2}} \right ).
\end{align}
The overall end-to-end communication in direct transmission occurs in two phases and hence there is a pre-log factor $1/2$ in the above expression. Thus, the OP for the direct transmission between two PUs is
\begin{align}
\mathcal{P}_{\textmd{out},\hat{j}}^{(\textmd{D})}=\Pr\left [\frac{P_{j}\left | h_{j,\hat{j}} \right |^{\textmd{2}}}{\sigma^{2}} <  \widetilde{\gamma_{\hat{j}}} \right] = \frac {\Upsilon \left [ m_{j\hat{j}}, \frac{m_{j\hat{j}} \widetilde{\gamma_{\hat{j}}} \sigma^{2}}{\Omega _{j\hat{j}} P_{j}}  \right ] }{\Gamma \left [ m_{j\hat{j}} \right ]}
\end{align}
where $\widetilde{\gamma_{\hat{j}}}=2^{2r_{\hat{j}}}-1$. For effective spectrum sharing, the SWIPT-enabled CCRN should lead to lower or equal OP than that of direct transmission (without relaying) under the same required target rate \cite{DSIoT}, i.e.,
\begin{align}
\mathcal{P}_{\textmd{out},a} \leq \mathcal{P}_{\textmd{out},a}^{(\textmd{D})} \text{ and }   \mathcal{P}_{\textmd{out},b} \leq \mathcal{P}_{\textmd{out},b}^{(\textmd{D})}.
\end{align}
 On the other hand, maintaining QoS of the primary system is also essential while enabling spectrum sharing for the secondary transmission. For maintaining QoS of the primary system, the term corresponding to OP of the direct transmission ($\mathcal{P}_{\textmd{out},b}^{(\textmd{D})}$) can be replaced by a predefined target OP to support the required target rate. Using the above expressions, the critical value of spectrum sharing factor $\mu^{\star}$, beyond which the proposed system achieves lower OP than direct transmission link, can be obtained using numerical methods. Therefore, the network can assure effective spectrum sharing for $\mu_{i} \geq \mu^{\star}$. Note that, the value of $\mu^{\star}$ is crucial to define the range of constraints for solving the optimization problem in \eqref{main_problem}. Thus, it has been ensured that the system under consideration satisfies the QoS constraint.

\subsection{System Throughput}
Under the delay-limited transmission mode of the considered scheme, throughput can be calculated as the sum of average target rates of both the primary and secondary systems that can be attained successfully over the fading channels \cite{LiuIET,DSIoT}. Thus, system throughput can be defined based on OPs of primary and secondary systems, each at a fixed transmission rate. Here, all the individual data rates are considered to be equal to the target transmission data rate $r_{\textmd{th}}$ in bits/sec/Hz. Therefore, the system throughput can be expressed as
 \begin{align}\label{st}
 \mathcal{S}_{\textmd{T}} = \mathcal{S}_{\textmd{PU}} + \mathcal{S}_{\textmd{SU}}
 \end{align}
  where $\mathcal{S}_{\textmd{PU}}$ and $\mathcal{S}_{\textmd{SU}}$ are the throughputs of primary and secondary users, respectively. Furthermore, $\mathcal{S}_{\textmd{PU}}$ and $\mathcal{S}_{\textmd{SU}}$ can be expressed in terms of their respective individual outage probabilities as
   \begin{align}
 \mathcal{S}_{\textmd{PU}} = \frac{\left (1-\alpha  \right ) }{3}\bigg( \left (1-\mathcal{P}_{\textmd{out},a}\right )r_{a}+\left ( 1-\mathcal{P}_{\textmd{out},b} \right )r_{b} \bigg)
 \end{align}
   \begin{align}
\mathcal{S}_{\textmd{SU}} = \frac{\left (1-\alpha  \right ) }{3}\bigg( \left (1-\mathcal{P}_{\textmd{out},1}\right )r_{1}+\left ( 1-\mathcal{P}_{\textmd{out},2} \right )r_{2} \bigg)
\end{align}
where the primary outage probabilities $\mathcal{P}_{\textmd{out},a}$ $\&$ $\mathcal{P}_{\textmd{out},b}$ are defined in \eqref{sjeo}, and secondary outage probabilities $\mathcal{P}_{\textmd{out},1}$ $\&$ $\mathcal{P}_{\textmd{out},2}$ are given in \eqref{oopd}.

\subsection{Energy Efficiency}\label{EE}
For any communication system, it is desirable to support an increased system throughput while minimizing energy consumption, and hence contributing to environment-friendly transmission. Thus, energy efficiency has become an important parameter in designing and analyzing the system performance. Energy efficiency is defined as \cite{LiuIET}
 \begin{align}
\textmd{Energy efficiency}= \frac{\textmd{Total amount of data delivered}}{\textmd{Total amount of energy consumed}}
 \end{align}
It follows that the energy efficiency for the considered system in the delay-limited scenario is given by
\begin{align}\label{ee}
\eta_{\rm EE}=\frac{\mathcal{S}_{\textmd{T}}}{\left(\alpha + \frac{\beta \left ( 1-\alpha \right )}{3}\right)\left ( P_{a}+P_{b} \right )}.
 \end{align}

\subsection{Optimization}\label{Optimization}
Given the above system throughput and energy efficiency expressions and variables, we now formulate an optimization problem with an objective of maximizing the system throughput and energy efficiency while satisfying the system parameters constraints. Here, we strive to find the optimal values of TS, PS and spectrum sharing factors, i.e., $\alpha$, $\beta$ and $\mu$ that maximize the system throughput. The formulated optimization problem is as follows:
\begin{align}
\mathbf{P1:}~~~ &\underset{\alpha, \beta, \mu}{\textrm{maximize~}}
\mathbb{U}(\alpha, \beta, \mu)  \label{main_problem} \\
\text{s.t.} \quad
& \textrm{(Time switching constraint):} \nonumber\\
& \text{C}_1:\; 0 \leq \alpha < 1 \nonumber\\
& \textrm{(Power splitting constraint):} \nonumber\\
& \text{C}_2:\; 0 \leq \beta < 1  \nonumber\\
& \textrm{(Spectrum sharing constraint):} \nonumber\\
& \text{C}_3:\; \mu^\star \leq \mu < 1, \nonumber
\end{align}
where $\mathbb{U}\in\{\mathcal{S}_{\rm T}, \eta_{\rm EE}\}$ is the objective function and the expressions of $\mathcal{S}_{\rm T}$ and  $\eta_{\rm EE}$ are provided in \eqref{st} and \eqref{ee}, respectively.

Similar to \cite{alsharoa2017optimization}, it can be seen that problem $\mathbf{P}_1$ is non-convex in nature, and finding an optimal solution is computationally challenging. Therefore, in this paper, the particle swarm optimization technique is implemented to maximize the system throughput, energy efficiency and thereby obtain near optimal values of TS, PS and spectrum sharing factors (i.e., $\alpha$, $\beta$ and $\mu$, respectively). PSO is a population-based stochastic optimization technique known for its ease of implementation, accuracy and robustness. The algorithm is initialized with $N$ number of random solutions called as population. Here, a solution includes the values of $\alpha$, $\beta$ and $\mu$. The system throughput and energy efficiency corresponding to a solution set are evaluated using \eqref{st} and \eqref{ee}. The algorithm updates the population for every iteration based on the individual solutions' best performance, called $p^{(\rm best)}$ or local best performance, and the overall best performance of the population set, called $g^{(\rm best)}$ or global best performance. Therefore, $p^{(\rm best)}$ implies the solutions' individual best performance which gives maximum system throughput and energy efficiency, whereas $g^{(\rm best)}$ implies the best solution in the entire population set which yields the maximum system throughput and energy efficiency. In this algorithm, the solutions are updated for every iteration using two parameters called \emph{velocity} and \emph{position} \cite{Shi}. The velocity of a solution implies the values by which $\alpha$, $\beta$ and $\mu$ have to be changed in order to update the solution. The position of a solution is the updated set of $[\alpha, \beta, \mu]$. Here, velocity has been evaluated separately for $\alpha$, $\beta$ and $\mu$. The velocity $v$ and position $\psi$ of the solution are updated using the following equations:
\begin{align}\label{velocity}
v^{(\textmd{next})}_{i} = w*v^{\textmd{(current)}}_{i}+c_{1}*\textmd{rand()}*(p^{(\rm best)}_{i}-\psi_{i})\nonumber\\
+c_{2}*\textmd{rand()}*(g^{(\rm best)}-\psi_{i})
\end{align}
\begin{align}\label{position}
 \psi^{(\textmd{next})}_{i}= \psi^{\textmd{(current)}}_{i}+v^{(\textmd{next})}_{i}.
 \end{align}
In the above expressions, $v_{i}$ is the velocity of the $i^{\textmd{th}}$ solution, $w$ is the inertia weight factor, $c_{1}$ and $c_{2}$ are learning factors, $\mathrm{rand}()$ is a uniformly distributed random number between 0 and 1, $p^{(\rm best)}_{i}$ is the individual best performance of the $i^{\textmd{th}}$ solution, $g^{(\rm best)}$ is the best solution of the population set and $\psi_{i}$ denotes the position of $i^{\textmd{th}}$ solution. If the updated position is found to be out of the search space, then its value is set to be within the bounds and the corresponding velocity of the solution is equated to zero. The termination criterion of the algorithm is given by the number of iterations.

The proposed PSO Algorithm 1 process in three steps: (i) initialization, (ii) update, and (iii) termination. Line 1 and 2 are initializations and contribute one operation for $N$ times each. Line 4 performs $N$ iterations to find system throughput and energy efficiency. The calculation of
	internal expressions in system throughput and energy efficiency is ignored as they can be saved in a lookup table. In line 5, using bubble sort, we find $p^\textrm{(best)}$ which requires $N\log(N)$ operations, and finding $g^\textrm{(best)}$ in $N$ solutions requires $\log(N)$ operations. In line 6, velocity update for N particles and three parameters require 15N multiplications and 12N additions. To update its values, one more addition is required. All the above operations are performed $t$ times to check the termination condition.
	Therefore, the worst-case complexity is $O(N\cdot t \cdot \log(N))$.

\begin{table}[h!]
	\centering
	\caption{Number of terms required in infinite series for achieving required accuracy.}
	\label{tab:my-table}
	\begin{tabular}{c|c|c|c|c}
		\hline\hline
		\multirow{2}{*}{Index $s$} & \multicolumn{2}{c|}{$P_{\text{out},a}$ in \eqref{oopc}} & \multicolumn{2}{c}{$P_{\text{out},1}$ in \eqref{oopd}} \\ \cline{2-5}
		& $\mathrm{SNR} = 15$dB & 20dB & 20dB & 25dB \\ \hline
		1 & 0.111445 & 0.0133768 & 0.330062 & 0.20458 \\ \hline
		2 & 0.111445 & 0.0133768  & 0.4865 & 0.292226 \\ \hline
		3 & 0.111445  & 0.0133768  & 0.608107 & 0.356532 \\ \hline
		4 & 0.111445  & 0.0133768  & 0.696492 & 0.400999 \\ \hline
		5 & 0.111445   & 0.0133768  & 0.758037 & 0.430666 \\ \hline
		6 & 0.111445  & 0.0133768 & 0.799646 & 0.450000 \\ \hline
		7 & 0.111445  & 0.0133768 & 0.827179  & 0.462397 \\ \hline
		8 & 0.111445  & 0.0133768  & 0.845107  & 0.470251 \\ \hline
		9 & 0.111445  & 0.0133768 & 0.856636 & 0.475184 \\ \hline
		10 & 0.111445  & 0.0133768  & 0.863977 & 0.478261 \\ \hline
		11 & 0.111445  & 0.0133768& 0.868615 & 0.48017  \\ \hline
		12 & 0.111445  & 0.0133768 & 0.871526 & 0.481349 \\ \hline
		13 & 0.111445  & 0.0133768 & 0.873344 & 0.482075 \\ \hline
		14 & 0.111445  & 0.0133768 & 0.874473 & 0.48252 \\ \hline
		15 & 0.111445  & 0.0133768 & 0.875173 & 0.482793 \\ \hline
		16 & 0.111445  & 0.0133768 & 0.875605 & 0.482959 \\ \hline\hline
	\end{tabular}
\end{table}

\begin{algorithm}
\SetKwInOut{Input}{Input}
\SetKwInOut{Output}{Output}
\SetKwFor{While}{while}{}{end while}
\SetAlgoLined
 \Input{($\alpha$, $\beta$ and $\mu$)}
\Output{Maximal system throughput, energy efficiency and optimal solution $[\alpha^*, \beta^*, \mu^*]$}
    Initialize all the required parameters and the range (search space) of values of $\alpha$, $\beta$ and $\mu$.

    Randomly initialize a population set of solutions ($[\alpha, \beta, \mu]$), with an initial random position and velocity values.

    \While{termination criterion not met}
    {
    	Evaluate system throughput and energy efficiency of every solution using \eqref{st} and \eqref{ee}.
    	
    	Find $p^{(\rm best)}$ of every solution and $g^{(\rm best)}$ solution of the population set.
    	
    	Update the position and velocity of all the solutions using \eqref{velocity} and \eqref{position}
    	
    	 \If{position of the solution is out of bounds}{
    	      Set the position equal to the bound.
    	
    	      Set the velocity of the corresponding solution equal to zero.
    	}
    	
    	Update the population set.
    	
    	Repeat the procedure.
    }

 	\caption{Algorithm for maximizing system throughput and energy efficiency}

 \end{algorithm}

\section{Numerical and Simulation Results}\label{Num}

\subsection{Outage Probability}
This section discusses the outage performance at both the primary and secondary users with respect to various system parameters. All the numerical and simulation results are obtained under the assumption that $P_{a}=P_{b}=P$, $\sigma^2_{i}=\sigma^2_{j}=\sigma^2_{c_{i}}=\sigma^2$ for $i\in\{1,2\}$ and $j\in \{a,b\}$, and the SNR at the transmitting end is defined as ${P}/{\sigma^{2}}$ \cite{Wang2016}. 

Furthermore, a 2-D network topology is considered where $\sf PU_{a}$ and $\sf PU_{b}$ are placed at coordinates (0, 0), (4, 0), and $\sf SU_{1}$ and $\sf SU_{2}$ are at $(d, 0)$, $(d, 2)$, respectively, where $d=2$. Regarding the path-loss model, the path-loss exponent is selected as $\nu=3$, while the channel coefficient is modeled in terms of the distance between the end nodes and the path-loss exponent. For links ${\sf PU}_{a}\rightarrow {\sf SU}_{1}$ and ${\sf PU}_{a}\rightarrow {\sf SU}_{2}$, variances of the channel gains are represented as $\Omega_{1a}=d_{1a}^{-\nu}$ and $\Omega_{2a}=d_{2a}^{-\nu}$, respectively. Likewise, for ${\sf PU}_{b}\rightarrow {\sf SU}_{1}$, ${\sf PU}_{b}\rightarrow {\sf SU}_{2}$, and ${\sf SU}_{2}\rightarrow {\sf SU}_{1}$ links, variances of the channel gains are $\Omega_{1b}=d_{1b}^{-\nu}$, $\Omega_{2b}=d_{2b}^{-\nu}$, and $\Omega_{12}=d_{12}^{-\nu}$, respectively \cite{SourabhTVT}, \cite{LiuIET}. Also, $P_{\text{th}}$ is set as $-10$dBm and the noise power at both primary and secondary users is assumed to be $-60$dBm, unless specified otherwise. The fading severity parameters are assumed as $m_{1a}=m_{2a}=m_{a}$ and $m_{1b}=m_{2b}=m_{b}$. The other parameters vary with each figure and are defined therein. Table I shows the required number of terms in infinite series to achieve a fair accuracy in evaluating \eqref{oopc} and \eqref{oopd} for the considered set of parameters.

\begin{figure}[t!]
	\centering
	\includegraphics[width=3.5in]{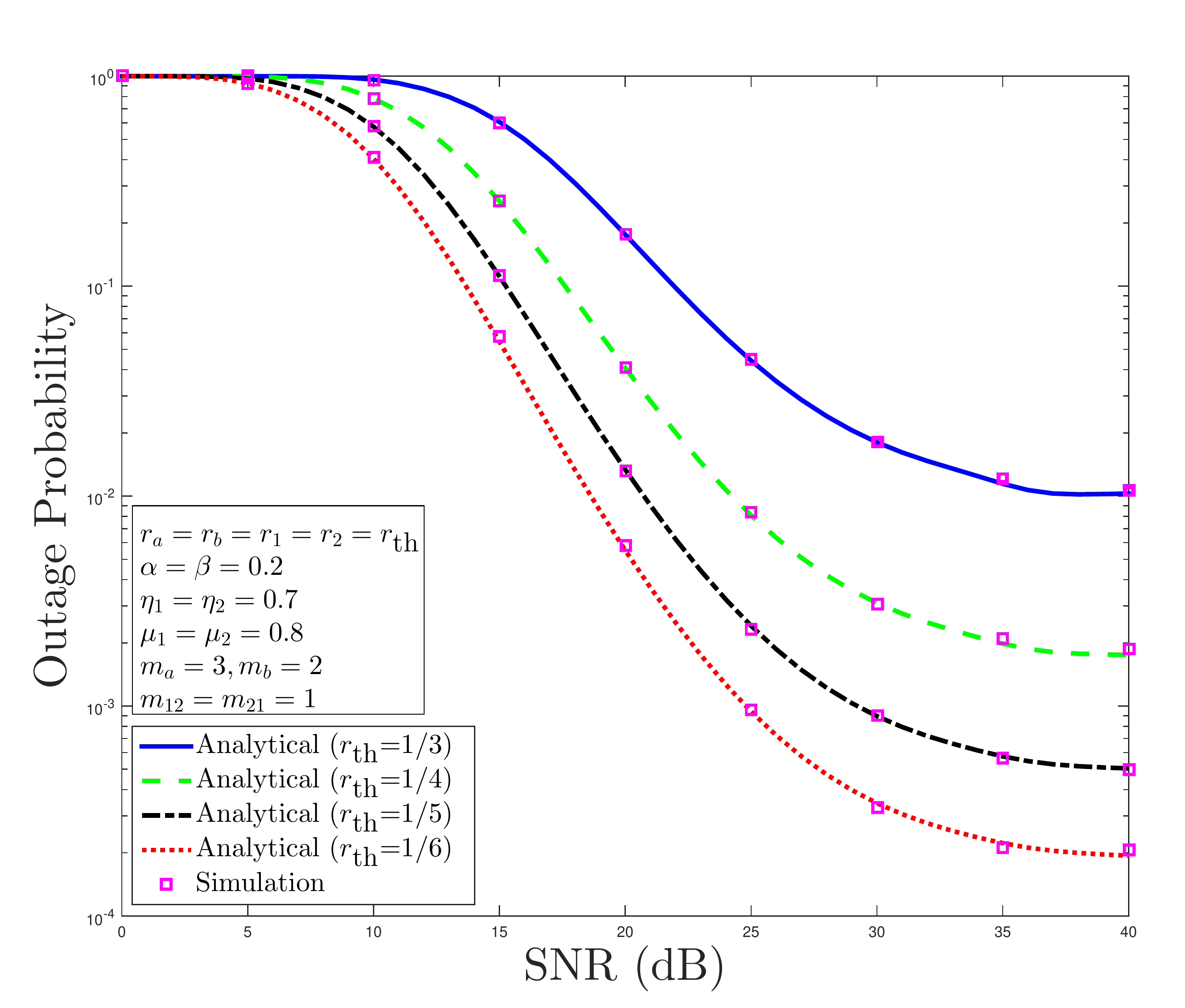}
	\caption{OP versus SNR for ${\sf PU}_{b}\rightarrow {\sf PU}_{a}$ link.}
	\label{figpusnr}
\end{figure}

\begin{figure}[t!]
	\centering
	\includegraphics[width=3.5in]{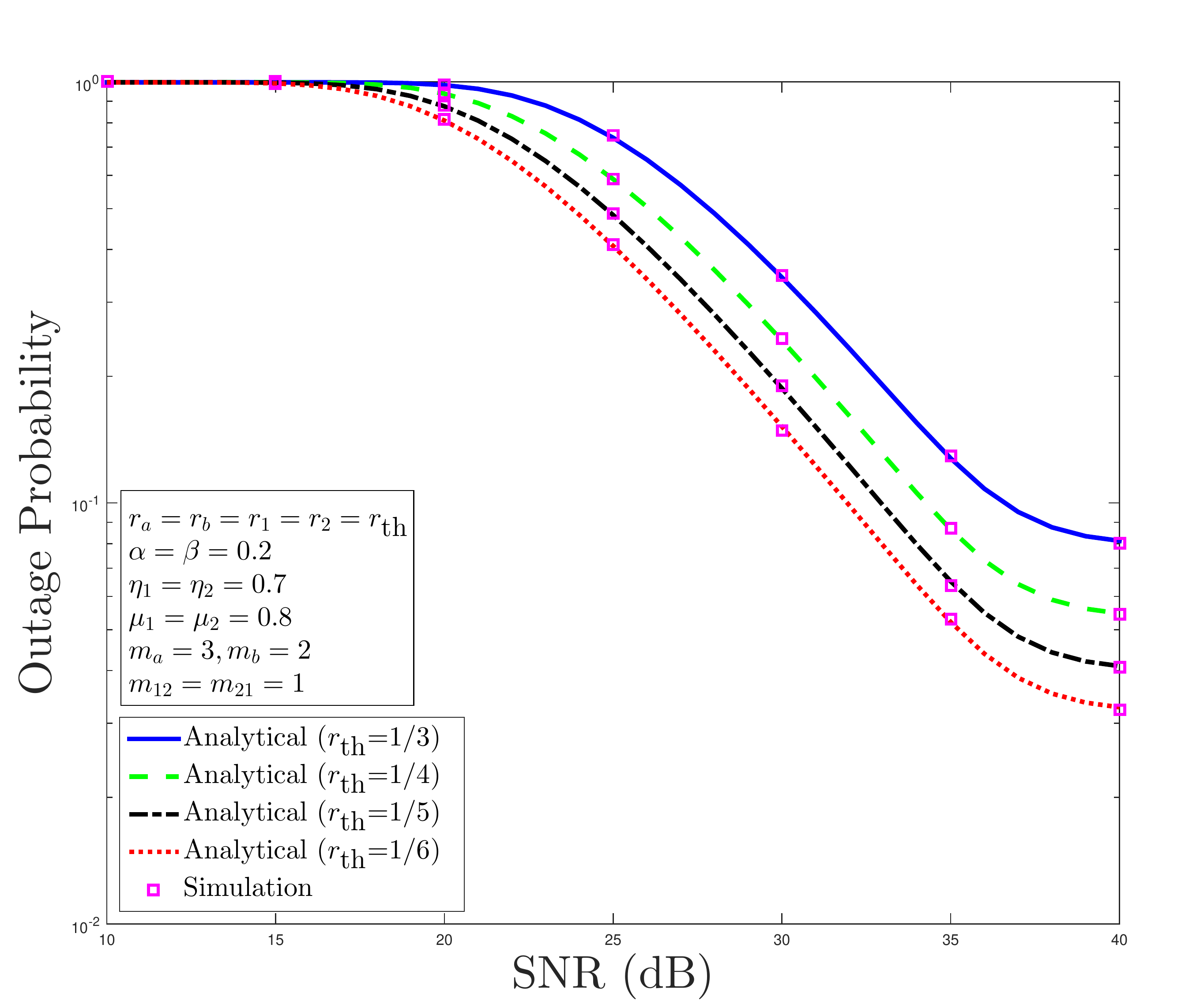}
	\caption{OP versus SNR for ${\sf SU}_{2}\rightarrow {\sf SU}_{1}$ link.}
	\label{figsusnr}
\end{figure}

Fig. \ref{figpusnr} shows the primary link outage probability-versus-SNR curves for different target data rates. In plotting this figure, the parameters are set as $\alpha=\beta=0.2$, $\mu=0.8$, $m_{a}=3$ and $m_{b}=2$. From this figure, one can notice that the OP performance improves with the increase in SNR values up to a certain value and remains constant afterward in the high SNR range. This is due to the fact that, since a non-linear EH model is used, the total transmit power becomes constant after it exceeds the threshold value $P_{\text{th}}$. Also, the OP performance of the primary link is better for lower target rates because lower target rates result in relatively lower target SNRs. Hence, the probability of outage occurrence becomes comparatively lower with lower values of target SNRs. Furthermore, Fig. \ref{figsusnr} plots the secondary link OP-versus-SNR curves for different target data rates. With the increase in SNR values, the behavior of OP performance at the SU is similar to that of the OP performance at PU. Here also, the OP curves become saturated at the high SNR range, which is attributed to the non-linearity of the considered energy harvester.

\begin{figure}[t!]
	\centering
	\includegraphics[width=3.5in]{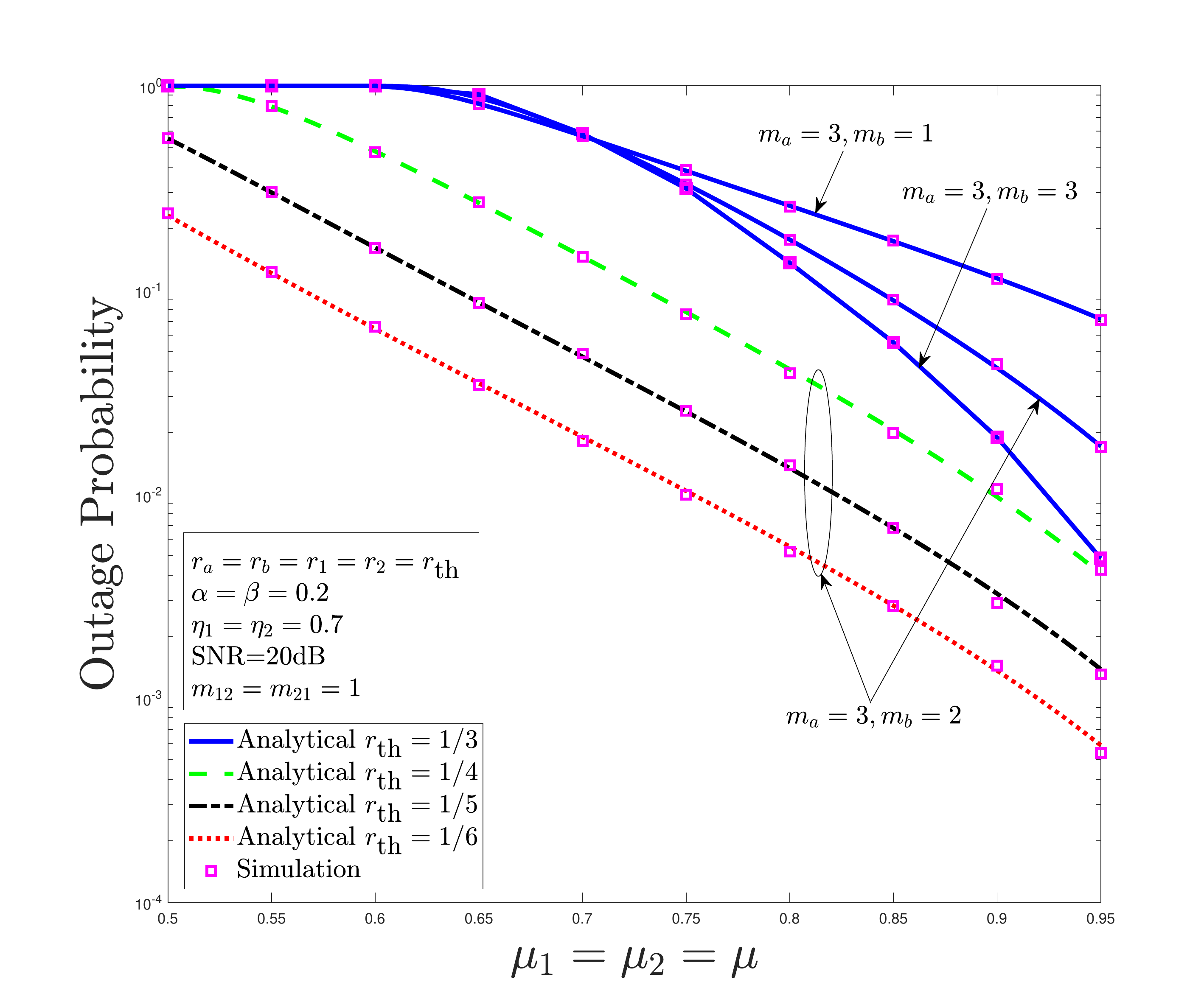}
	\caption{OP versus $\mu$ for ${\sf PU}_{b}\rightarrow {\sf PU}_{a}$ link.}
	\label{figpmu}
\end{figure}

\begin{figure}[t!]
	\centering
	\includegraphics[width=3.5in]{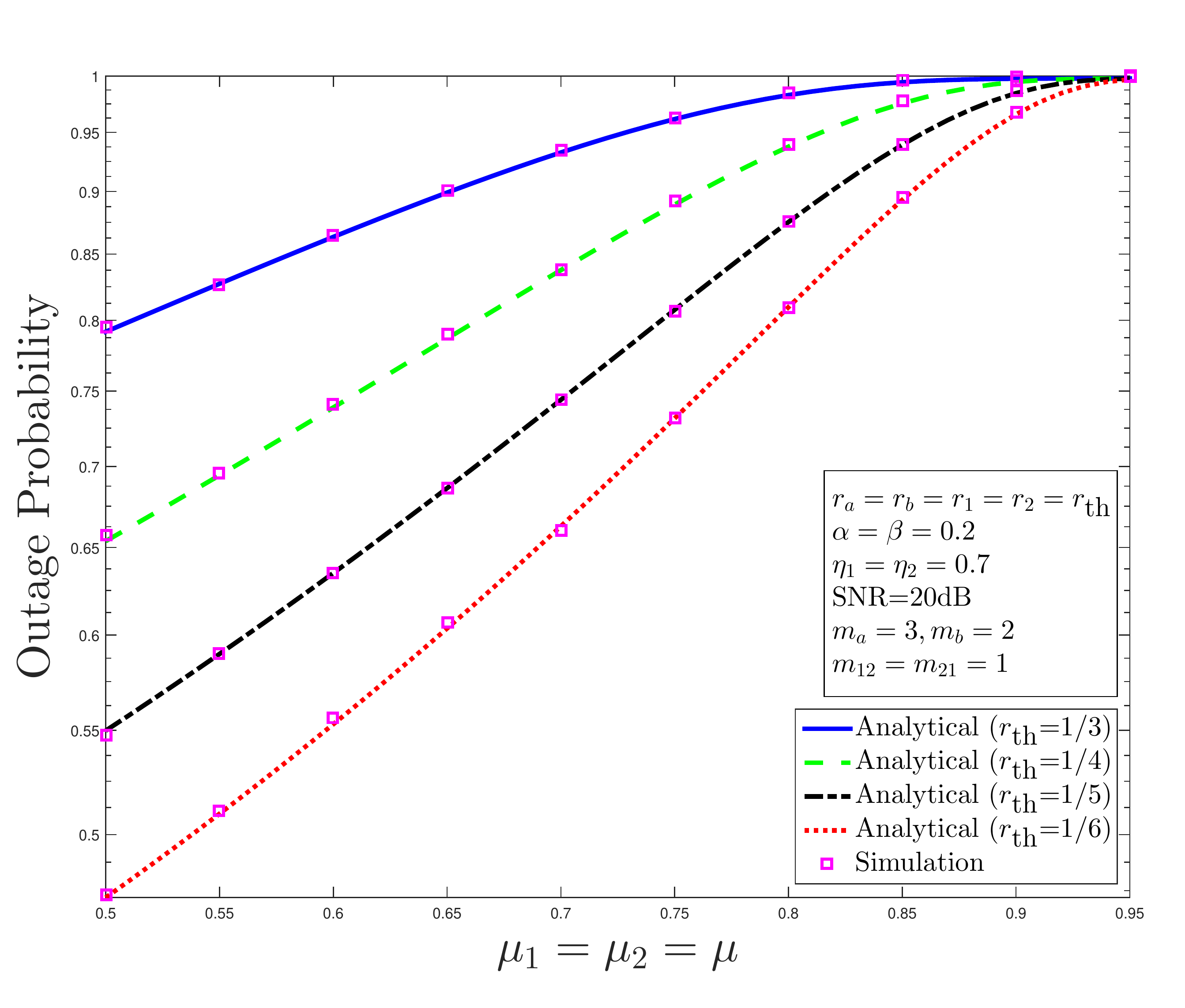}
	\caption{OP versus $\mu$ for ${\sf SU}_{2}\rightarrow {\sf SU}_{1}$ link.}
	\label{figsmu}
\end{figure}

\begin{figure}[t!]
	\centering
	\includegraphics[width=3.7in]{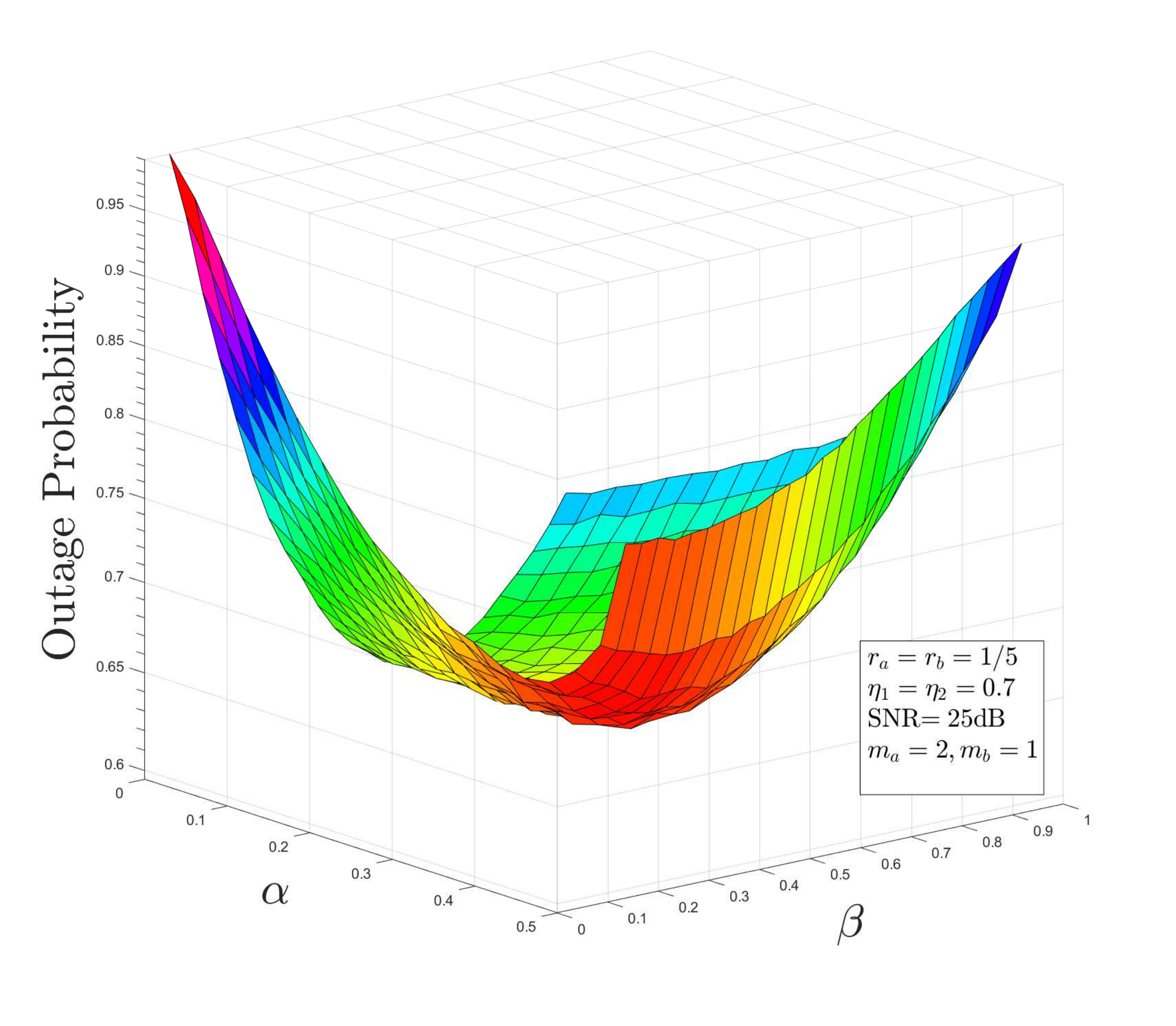}
	\caption{OP versus $\alpha$ and $\beta$ for ${\sf PU}_{b}\rightarrow {\sf PU}_{a}$ link.}
	\label{fig4}
\end{figure}

\subsection{OP for PUs and SUs with Spectrum Sharing Factor}
For Fig. \ref{figpmu}, the values of different parameters are considered as $\eta_{1}=\eta_{2}=0.7$, $\alpha=0.2$, $\beta=0.2$, $\mu_{1}=\mu_{2}=\mu$ and $\textmd{SNR}=20$dB. This figure shows the OP of the primary link ${\sf PU}_{b}\rightarrow {\sf PU}_{a}$ versus spectrum sharing factor $\mu$, for various target rates. On analyzing the OP performance of the primary system for varying fading severity parameters, it is observed that the OP performance improves with increasing value of $m_{b}$. This is in accordance with the fact that, when the system operates in a less severe fading environment, it has better link reliability. Note that for a specific target data rate, the value of $\mu_{i}$ should be greater than a certain value, in order to enable spectrum sharing. For $r_{j}=1/3$ bps/Hz and $\alpha=0.2$, the acceptable spectrum sharing factor range is $0.62<\mu_{i}<1$, while for $\mu_{i}<0.62$ outage occurs at the primary system (i.e., the OP equals one). Along with the target rate required at the node, the TS factor $\alpha$ also determines the minimum possible value of $\mu_{i}$. Further, with the increase in the value of $\mu_{i}$, the primary link OP also improves due to the fact that a larger portion of power is being allotted for primary transmissions.

Fig. \ref{figsmu} shows the OP curves for the link ${\sf SU}_{2}\rightarrow {\sf SU}_{1}$ versus $\mu$, under different fading scenarios and target rates. All the parameters considered here are the same as in Fig. \ref{figpmu}. In contrast to Fig. \ref{figpmu}, as the value of $\mu_{i}$ increases, the secondary link OP performance deteriorates. This is because the transmitted power at the secondary system is scaled by a factor of $(1-\mu_{i})$. Here also, it can observed that when the fading severity parameter $m_{12}$ increases, the secondary system shows better performance due to the fact that link reliability is better in a less severe fading environment. Unlike the primary link, the secondary nodes can have reliable communication for the complete range of $\mu_{i}$. Likewise, the OP performance of the secondary system also degrades with increasing target rate. All the simulation results and analytical plots of Figs. \ref{figpmu} and \ref{figsmu} are in good consonance, which confirms the accuracy of the theoretical analysis.

\subsection{OP for PUs with TS and PS Factors}

In the considered CCRN, hybrid SWIPT is employed and therefore it is of particular interest to investigate the joint effect of both TS and PS factors on the OP performance. Here, the noise power at PUs is set as $-10$dBm. The OP curves are plotted against TS factor $\alpha$ and PS factor $\beta$ for the primary system as shown in Fig. \ref{fig4}. For the considered parameters, the primary link achieves the minimum OP when $\alpha=0.08$ and $\beta=0.75$. In contrast, when $\alpha=0$, the primary link attains the minimum OP at $\beta=0.8$. For a certain $\alpha$ value and increasing $\beta$ value, the OP of the primary link decreases up to a certain point and then starts increasing. Also, it can be observed that for certain $\alpha$ and $\beta$ values, the OP is very low compared to that at other values.

\subsection{Effective Spectrum Sharing}

\begin{figure}[t!]
	\centering
	\includegraphics[width=3.5in]{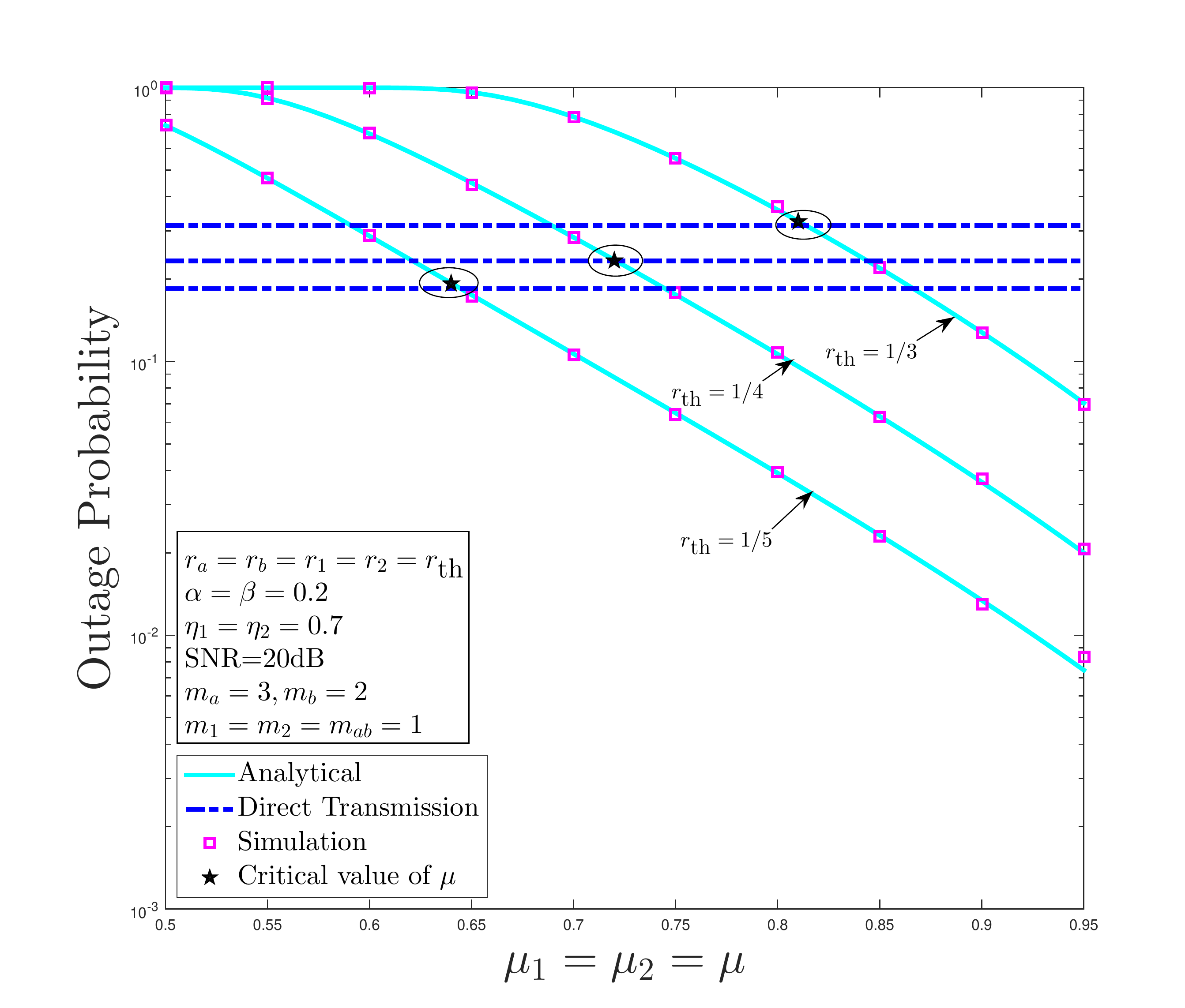}
	\caption{OP versus $\mu$ for ${\sf PU}_{b}\rightarrow {\sf PU}_{a}$ link with proposed scheme and direct transmissions.}
	\label{figmucritical}
\end{figure}

Fig. \ref{figmucritical} plots OP curves for the primary link ${\sf PU}_{b}\rightarrow {\sf PU}_{a}$ versus the spectrum sharing factor $\mu$. This figure demonstrates the outage performance of both the proposed scheme and direct transmission of the primary link and offers some key insights into the system design constraints for effective spectrum sharing with respect to $\mu$. Here, the parameters are considered as $\alpha=0.2$, $\beta=0.2$ and $\textmd{SNR}=20$dB. It can be observed from the figure that the critical value of $\mu_{i}$, denoted as $\mu^{\star}$, is the point at which the OPs of the proposed system and that of the direct transmission  are equal. For $\mu_{i} > \mu^{\star}$, the OP of the primary system shows better performance than the direct transmission link. Therefore, $\mu^{\star}<\mu_{i}<1$ can be considered as the feasible range of $\mu_{i}$ for effective spectrum sharing. Below the critical value $\mu^{\star}$, the system shows poor outage performance compared to direct primary transmission link. It is also noted that, with increasing data rates, the value of $\mu^{\star}$ also increases. This is in agreement with the fact that, as higher data rates result in higher target SNRs, the fraction of power allocated for primary signal transmission in BC phases should be greater for higher data rates. Consequently, the critical value of $\mu_{i}$, which is the power splitting factor for spectrum sharing, also becomes larger.

\subsection{System Throughput}
\begin{figure}[t!]
	\centering
	\includegraphics[width=3.5in]{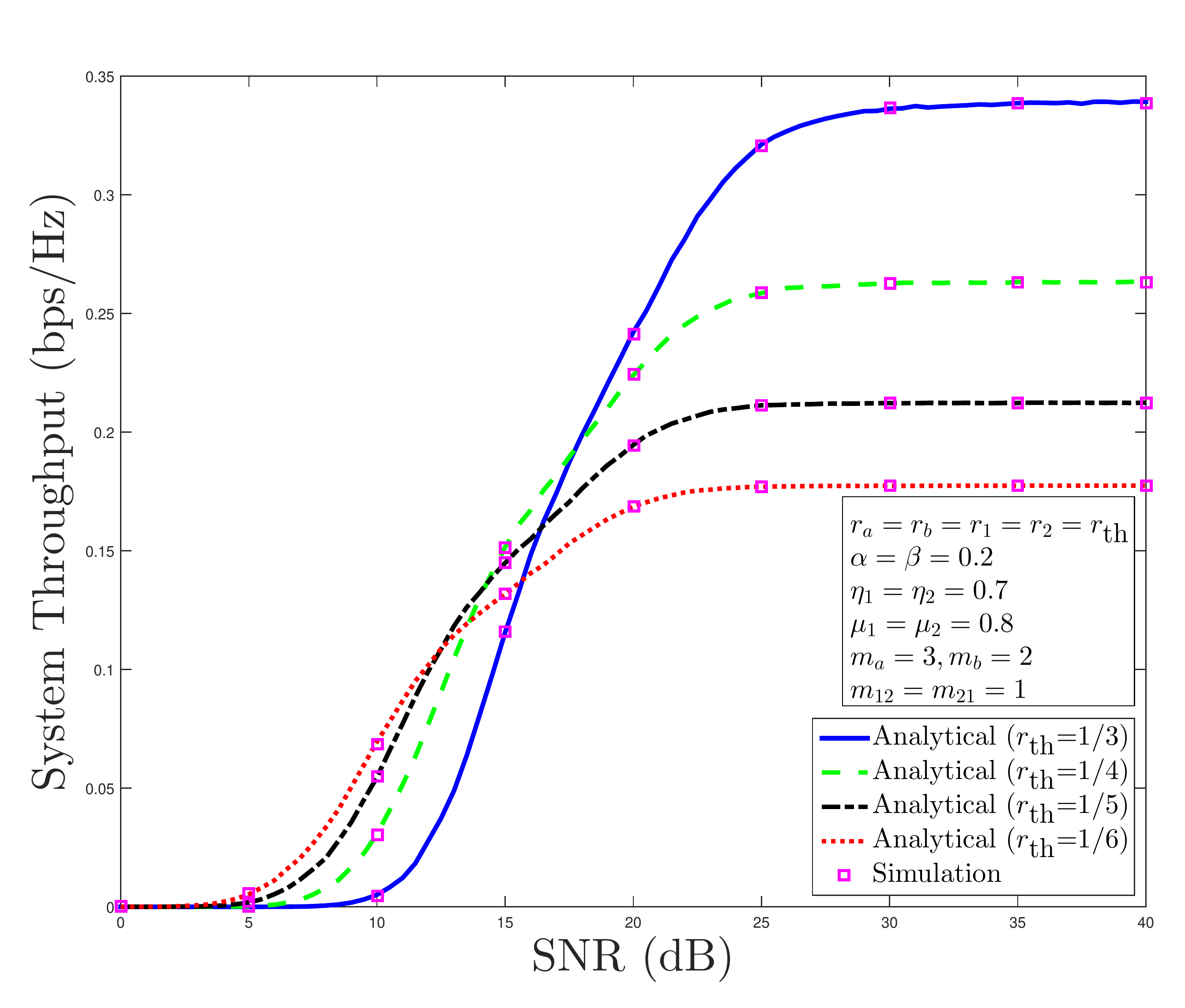}
	\caption{System throughput versus SNR for different target rates.}
	\label{figstsnr}
\end{figure}


Fig. \ref{figstsnr} shows the system throughput in bps/Hz versus SNR in dB. System throughput curves are plotted for different values of target rate $r_\textmd{th}$. The parameters are considered as $\alpha= \beta = 0.2$, $\mu= 0.8$, $\eta_1=\eta_2 = 0.7$, $m_a=3$, $m_b=2$, $m_{12} = m_{21} = 1$ and $r_a = r_b = r_1 = r_2 = r_\textmd{th}$. The target rates of primary and secondary users are assumed to be equal. It can be observed that, at the low SNR range, the system throughput corresponding to higher target rates is very small compared to that at lower target rates. This is attributed to the fact that, as the target rates increase, the target SNR also increases, but because the considered SNR is low, the outage probabilities at both PUs and SUs increase, which results in decreased system throughput values. When the SNR is in the medium to high range, the outage probabilities improve with increasing target rates and hence the system throughput also increases with the target rates. Also, it is observed that after a certain SNR value (which varies with target rates), the system throughput becomes saturated and serves as the maximum achievable throughput for the defined parameter set.

\subsection{Energy Efficiency}
\begin{figure}[t!]
	\centering
	\includegraphics[width=3.3in]{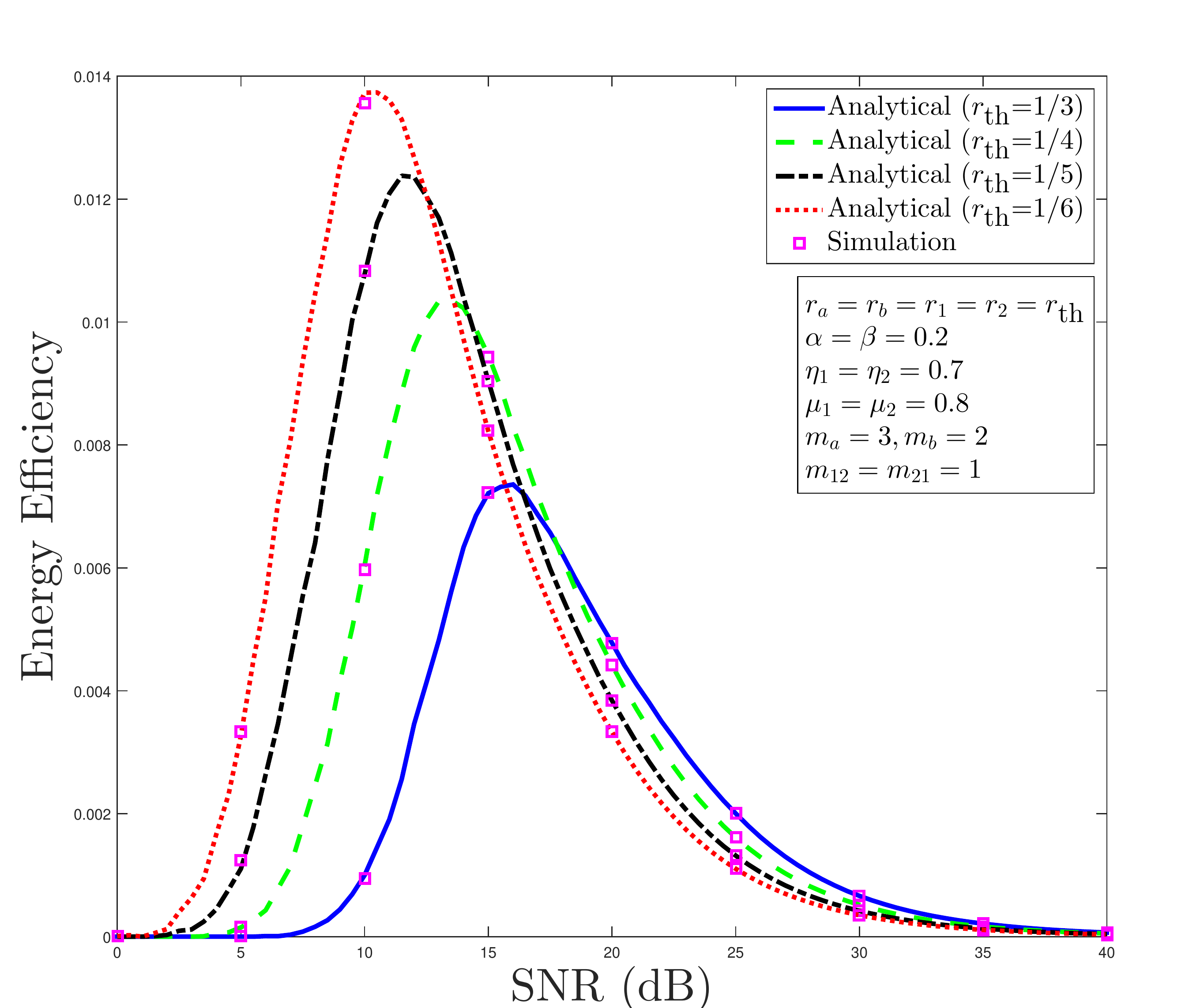}
	\caption{Energy efficiency versus SNR for different target rates.}
	\label{figeesnr}
\end{figure}

In Fig. \ref{figeesnr}, the energy efficiency of the system versus SNR is studied under different target rates. The system parameters considered are the same as that considered for Fig. \ref{figstsnr}. It is evident from the figure that energy efficiency of the system corresponding to a certain target rate is high only for a specific range of SNR values, and this range of SNR varies for different target rates. For lower target rates, maximum energy efficiency is achieved at lower SNR values and for higher target rates, maximum energy efficiency is attained at higher values of SNR. Also, as the target rates increase, the maximum attainable energy efficiency decreases. Thus, it can be deduced that the maximum energy efficiency for a certain target rate can be obtained only for a specific range of SNR values.

\subsection{Results with Optimized System Parameters}
\begin{figure}[t!]
	\centering
	\includegraphics[width=3.5in]{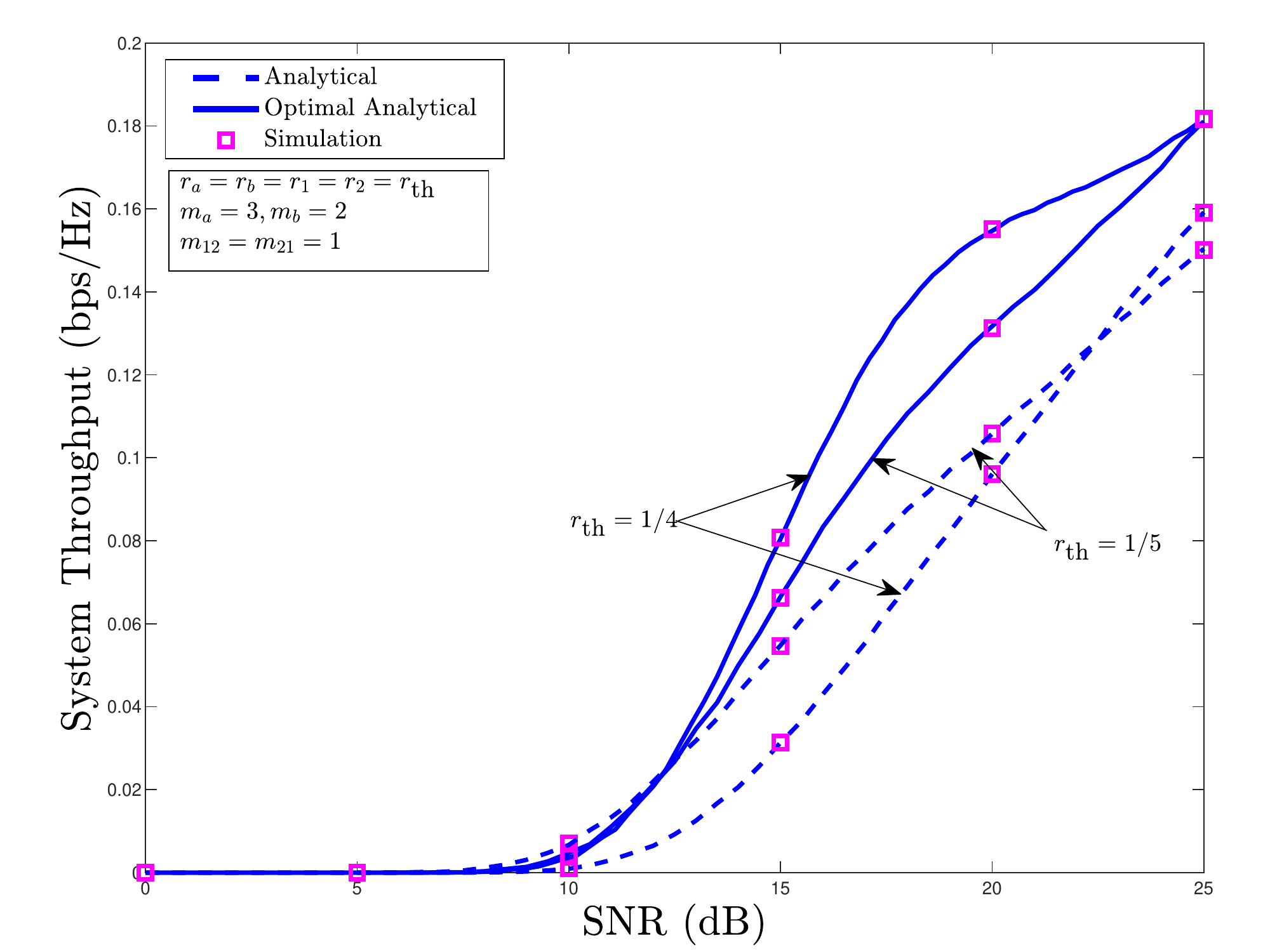}
	\caption{System throughput versus SNR for optimized and typical input parameters with different target rates.}
	\label{figstopt}
\end{figure}

\begin{figure}[h]
	\centering
	\includegraphics[width=3.5in]{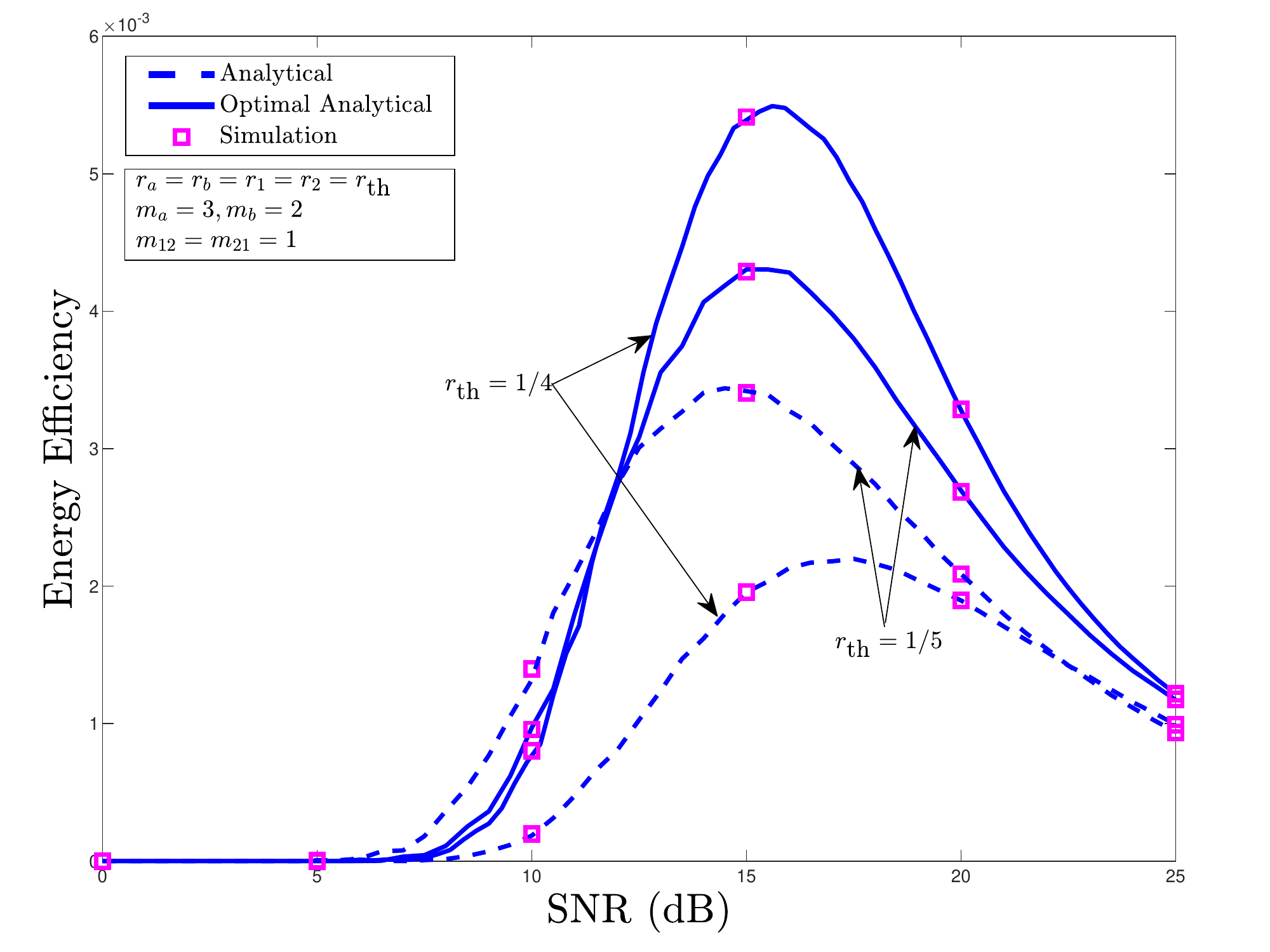}
	\caption{Energy efficiency  versus SNR for optimized and typical input parameters with different target rates.}
	\label{figeeopt}
\end{figure}

Figs. \ref{figstopt} and \ref{figeeopt} plot the system throughput and energy efficiency, respectively, for varying SNR. In Fig. \ref{figstopt} the curves are plotted for two sets of $\alpha$, $\beta$ and $\mu$ and for different target data rates. One set comprises of typical values as considered in previous figures whereas the other is the optimized set obtained from the optimization algorithm described in Section \ref{Optimization}. It can be observed that the system throughput corresponding to the optimized set of parameters shows a much superior performance compared to the curves corresponding to typical parameters. The difference in between the optimized system throughput and the typical system throughput is much higher in the medium to high SNR range as compared to the low SNR range because the optimal parameters are obtained for SNR $=20$ dB. Since the system throughput depends on the TS factor $\alpha$ and OP, and the optimization algorithm results in minimized OP and low $\alpha$ value, the resultant optimized system throughput is comparatively high in the high SNR range. Whereas in the low SNR region, outage occurs due to the higher target SNRs and thus the system throughput is zero for the low SNR range. After a saturation point, the system throughput remains constant at the maximum achievable throughput.

\begin{table}[h]
	\centering
	\caption{Optimal system parameters}
	\begin{tabular}{l|l|l|l}
		\cline{1-4}
		\multirow{2}{*}{Target rate} & \multicolumn{3}{l}{Optimal system parameters}           \\
		\cline{2-4}
		& $\alpha$ & $\beta$ & $\mu$       \\
		\hline
		1/2                          & 0.01                  & 0.6918               & 0.95  \\
		\hline
		1/3                          & 0.01                  & 0.7125              & 0.95  \\
		\hline
		1/4                          & 0.01                  & 0.6847              & 0.8165   \\
		\hline
		1/5                          & 0.01                  & 0.7101              & 0.7544  \\
		\hline
		1/6                          & 0.01                  & 0.7140             &  0.7732    \\
		\hline
	\end{tabular}
\end{table}

The optimal values of $\alpha, \beta$, and $\mu$, to plot curves in Fig. \ref{figstopt}, are given in Table II for SNR $=20$ dB. From this table, one can see that the values of $\alpha$ are close to zero. This is because of a well known fact that PS-based SWIPT shows better system throughput than TS-based SWIPT in the high SNR region \cite{atta}.

Similarly, energy efficiency curves are plotted in Fig. \ref{figeeopt} for specific target data rates and for typical and optimized sets of $\alpha$, $\beta$ and $\mu$. There is a significant difference between the optimal and typical energy efficiencies in the medium SNR range. This is again due to the fact that the optimized values of $\alpha$ and $\beta$ are very low compared to the typical values. Therefore in the mid SNR range the energy efficiency is very high compared to its typical counterpart. In the high SNR range, i.e., after the system attains the maximum throughput, the energy efficiency degrades with increasing SNR because as the SNR becomes high, the transmitted power, which is assumed to be proportional to the system SNR up to a certain threshold value, also increases and hence the energy efficiency decreases. Furthermore, with increasing SNR, the energy efficiency remains constant due to the non-linearity of the EH receiver. Thus, the system can achieve maximum throughput and energy efficiency for optimized values of $\alpha$, $\beta$ and $\mu$ at particular SNR value and target data rate.

\section{Conclusions}\label{Conclusion}
This paper has investigated a new hybrid SWIPT-based spectrum sharing system to enable two-way communications of PUs and SUs over the same licensed spectrum. A pair of AF-based SWIPT-enabled SUs provides cooperative relaying for primary signal transmissions. The primary signals are relayed by the SUs in two successive phases, thereby improving the reliability of the communication links. Outage performance of both primary and secondary systems with the considered scheme operating under the Nakagami-$m$ fading environment is analyzed. Further, the feasible range of spectrum sharing factor $\mu$ is deduced, for which the OP performance of the proposed scheme is superior to that of the direct transmission. The system throughput and energy efficiency expressions are formulated and they are maximized by means of particle swarm optimization. Numerical and simulation results confirmed the accuracy of the derived analytical expressions. The impact of crucial design parameters, e.g., spectrum sharing factor, TS and PS factors is revealed on performance of the considered system.


\appendices
\section{}\label{appA}
 Let $X_{i}\triangleq|h_{j,i}|^{2}$ and $Y_{i}\triangleq|h_{\hat{j},i}|^{2}$ for $i\in\{1,2\}$, $j,\hat{j}\in\{a,b\}$ with $j\neq\hat{j}$, be Gamma distributed random variables. Their PDFs can be given as  $f_{X_{i}}(x_{i})=\left(\frac{m_{ij}}{\Omega_{ij}}\right)^{m_{ij}} \frac{x^{m_{ij}-1}_{i}}{\Gamma[m_{ij}]}\mathrm{e}^{-\frac{m_{ij}x_{i}}{\Omega_{ij}}},\,x_{i}\geq0$, and $f_{Y_{i}}(y_{i})=\left(\frac{m_{i\hat{j}}}{\Omega_{i\hat{j}}}\right)^{m_{i\hat{j}}} \frac{y^{m_{i\hat{j}}-1}_{i}}{\Gamma[m_{i\hat{j}}]}\mathrm{e}^{-\frac{m_{i\hat{j}}y_{i}}{\Omega_{i\hat{j}}}},\,y_{i}\geq0$.
 The CDF $F_{\gamma_{i,j}^{(\text{lin})}}(\gamma_{j})$ can be formulated using (\ref{sdaet}), as
 \begin{align}\label{luuhh}
 F_{\gamma_{i,j}^{(\text{lin})}}(\gamma_{j})&=\textmd{Pr}\Bigg[\frac{\mu_{i}\Delta_{i} P_{\hat{j}}X_{i}Y_{i}}{\varepsilon_{ij}X_{i}+\omega_{ij}X_{i}^{2}+\omega_{i\hat{j}}X_{i}Y_{i}+\sigma^{2}_{j}}<\gamma_{j},\nonumber\\
  &  \quad \quad \quad \quad  \quad  \quad \quad \quad \quad \quad \quad \,\,\,\,\, P_{j}X_{i}+P_{\hat{j}}Y_{i} \leq P_{\text{th}}\Bigg]\nonumber\\
 &=\textmd{Pr}\left[Y_{i}<\frac{\varepsilon_{ij}\gamma_{j}+\omega_{ij}\gamma_{j}X_{i}}{\Xi}, Y_{i}\leq \frac{P_\text{th}-P_{j} X_{i}}{P_{\hat{j}}} \right]
 \end{align}
 where $\Xi=\mu_{i}\Delta_{i} P_{\hat{j}}-\omega_{i\hat{j}}\gamma_{j}$. From \eqref{luuhh}, one can observe that when the term $\Xi\leq0$, i.e., $\left(\mu_{i}-(1-\mu_{i})\gamma_{j}\right)\leq0$, the CDF $F_{\gamma_{i,j}^{(\text{lin})}}(\gamma_{j})=1$. On the other hand, when $\left(\mu_{i}-(1-\mu_{i})\gamma_{j}\right)>0$, the expression of $F_{\gamma_{i,j}^\text{lin}}(\gamma_{j})$ can be given as
 \begin{align}\label{ljljio}
 F_{\gamma_{i,j}^\text{lin}}(\gamma_{j})&=
 \int^{\iota_1}_{x_{i}=0}\!\!\!f_{X_{i}}(x_{i})
 \!\!\int^{\frac{\varepsilon_{ij}\gamma_{j}+\omega_{ij}\gamma_{j}x_{i}}{\Xi}}_{y_{i}=0}\!\!f_{Y_{i}}\!(\!y_{i}\!)dy_{i}\,dx_{i} \nonumber\\
 &+ \int^{\frac{P_\text{th}}{P_{j}}}_{x_{i}=\iota_1}\!\!\!f_{X_{i}}(x_{i})
 \!\!\int^{\frac{P_\text{th}-P_{j} x_{i}}{P_{\hat{j}}}}_{y_{i}=0}\!\!f_{Y_{i}}\!(\!y_{i}\!)dy_{i}\,dx_{i}
 \end{align}

 The CDF $F_{\gamma_{i,j}^{(\text{sat})}}(\gamma_{j})$ can be formulated using (\ref{sdaet_sat}), as,
 \begin{align}\label{luuhh_sat}
 F_{\gamma_{i,j}^{(\text{sat})}}(\gamma_{j})&=\textmd{Pr}\left[\frac{\mu_{i}\Delta_{i}P_{\text{th}}P_{\hat{j}}X_{i}Y_{i}} {\phi_{ij}X_{i}+\Phi_{ij}X_{i}^{2}+\Phi_{i\hat{j}}X_{i}Y_{i}^{2}+
 	\varphi_{\hat{j}j} Y_{i}}<\gamma_{j},\right.\nonumber\\
 & \left.\qquad\qquad \qquad\qquad\qquad\qquad P_{j}X_{i}+P_{\hat{j}}Y_{i} > P_{\text{th}}\right]\nonumber\\
 &=\textmd{Pr}\left[X_{i}>\frac{Y_{i}-\mathcal{T}_1}{\mathcal{T}_2}, X_{i}> \frac{P_\text{th}-P_{\hat{j}} Y_{i}}{P_{j}} \right].
 \end{align}
  where $\mathcal{T}_1 = \frac{\phi_{ij} \gamma_{j}}{\mu_{i}\Delta_{i}P_{\text{th}}P_{\hat{j}}-\Phi_{i\hat{j}} \gamma_{j}}$ and $\mathcal{T}_2 = \frac{\Phi_{ij} \gamma_{j}}{\mu_{i}\Delta_{i}P_{\text{th}}P_{\hat{j}}-\Phi_{i\hat{j}} \gamma_{j}}$.
 \begin{align}\label{ljljio_sat}
 F_{\gamma_{i,j}^\text{sat}}(\gamma_{j})&=
 \int^{\iota_2}_{y_{i}=0}\!\!\!f_{Y_{i}}(y_{i})
 \!\!\int^{\infty}_{x_{i}=\frac{P_\text{th}-P_{\hat{j}} y_{i}}{P_{j}}} \!\!f_{X_{i}}\!(\!x_{i}\!)dx_{i}\,dy_{i} \nonumber\\
 &+ \int^{\infty}_{y_{i}=\iota_2}\!\!\!f_{Y_{i}}(y_{i})
 \!\!\int^{\infty}_{x_{i}=\frac{y_{i}-\mathcal{T}_1}{\mathcal{T}_2}} \!\!f_{X_{i}}\!(\!x_{i}\!)dx_{i}\,dy_{i}.
 \end{align}
After substituting the PDFs in \eqref{ljljio} and \eqref{ljljio_sat} and applying \cite[eqs. 3.381.1, 3.471.9]{math}, the required solutions can be given as in Lemma \ref{lem1}.

\section{}\label{appB}
 Let $Z\triangleq|h_{i,\hat{i}}|^{2}$ be a random variable with PDF $f_{Z}(z)=\left(\frac{m_{i\hat{i}}}{\Omega_{i\hat{i}}}\right)^{m_{i\hat{i}}} \frac{z^{m_{i\hat{i}}-1}}{\Gamma[m_{i\hat{i}}]}\mathrm{e}^{-\frac{{m_{i\hat{i}}}z}{\Omega_{i\hat{i}}}},\,z\geq0$, and $W\triangleq P_{j}X_{i}+P_{\hat{j}}Y_{i}$ be a random variable whose PDF can be expressed as,
 \begin{align}
 \!\!\!f_{W}(w)&\!\!=\!\!\frac{\left(\frac{m_{ij}}{\Omega_{ij} P_{j}}\right)^{m_{ij}} \left(\frac{m_{i\hat{j}}}{\Omega_{i\hat{j}}P_{\hat{j}}}\right)^{m_{i\hat{j}}}}{\Gamma [m_{ij}] \Gamma [m_{i\hat{j}}]}\!\!\sum _{k=0}^{m_{ij}\!-\!1} (\!-\!1)^{-k\!+\!m_{ij}\!-\!1} \binom{m_{ij}\!-\!1}{k} \nonumber\\
 &\times \sum _{s=0}^{\infty} \frac{\left(\frac{m_{ij}}{\Omega_{ij} P_{j}}\right)^s}{s!} \left(\frac{m_{i\hat{j}}}{\Omega_{i\hat{j}} P_{\hat{j}}}\right)^{-(-k+m_{ij}+m_{i\hat{j}}+s-1)} w^k\nonumber\\
 &\times \mathrm{e}^{-\frac{m_{ij}}{\Omega_{ij} P_{j}} w} \Upsilon\left[-k+m_{ij}+m_{i\hat{j}}+s-1,\frac{m_{i\hat{j}}}{\Omega_{i\hat{j}} P_{\hat{j}}} w\right].
 \end{align}
From  \eqref{sadkdd}, the CDF $F_{\gamma_{i,\hat{i}}^\text{lin}}(\gamma_{\hat{i}})$ can be formulated as
 \begin{align}\label{jjjsg}
 F_{\gamma_{i,\hat{i}}^\text{lin}}(\gamma_{\hat{i}})&=\textmd{Pr}\left[\frac{\zeta_{i}W Z}{\xi_{i}Z+\sigma^{2}_{\hat{i}}}<\gamma_{\hat{i}}, W \leq P_{\text{th}}\right]\nonumber\\
 &=\textmd{Pr}\left[W< \frac{\xi_{i}\gamma_{\hat{i}}Z+\gamma_{\hat{i}}\sigma_{\hat{i}}^2}{\zeta_{i}Z}, W\leq P_\text{th}\right].
 \end{align}
 Following \eqref{jjjsg}, the CDF $F_{\gamma_{i,\hat{i}}^\text{lin}}({\gamma_{\hat{i}}})$ can be expressed in integration form as

 \begin{align}\label{kmklw}
 F_{\gamma_{i,\hat{i}}^\text{lin}}(\gamma_{\hat{i}})&= \int^{\iota_3}_{z=0}\!\!\!f_{Z}(z)
\!\!\int^{P_\text{th}}_{w=0} \!\!f_{W}\!(\!w\!)dw\,dz \nonumber\\
&+ \int^{\infty}_{z=\iota_3}\!\!\!f_{Z}(z)
\!\!\int^{\frac{\xi_{i}\gamma_{\hat{i}}z+\gamma_{\hat{i}}\sigma_{\hat{i}}^2}{\zeta_{i}z}}_{w=0} \!\!f_{W}\!(\!w\!)dw\,dz.
 \end{align}
The CDF of $F_{\gamma_{i,\hat{i}}^\text{sat}}(\gamma_{\hat{i}})$ can be formulated using  \eqref{sadkdd_sat} as
\begin{align}\label{jjjsg_sat}
 F_{\gamma_{i,\hat{i}}^\text{sat}}(\gamma_{\hat{i}})&=\textmd{Pr}\left[\frac{\Psi_{i}WZ}{\mathcal{C}_{i}|Z+\mathcal{D}_{\hat{i}}W} <\gamma_{\hat{i}}, W > P_{\text{th}}\right] \nonumber\\
 &=\textmd{Pr}\left[W < \frac{\mathcal{C}_{i} \gamma_{\hat{i}}Z}{\Psi_{i} Z-\mathcal{D}_{\hat{i}}\gamma_{\hat{i}} }, W > P_\text{th}\right] \nonumber\\
 &=\textmd{Pr}\left[W < \frac{\mathcal{C}_{i} \gamma_{\hat{i}}Z}{\Psi_{i} Z-\mathcal{D}_{\hat{i}}\gamma_{\hat{i}} }, W > P_\text{th}, Z > \frac{\mathcal{D}_{\hat{i}}\gamma_{\hat{i}}}{\Psi_{i}} \right] \nonumber\\
 & + \textmd{Pr}\left[W > \frac{\mathcal{C}_{i} \gamma_{\hat{i}}Z}{\Psi_{i} Z-\mathcal{D}_{\hat{i}}\gamma_{\hat{i}} }, W > P_\text{th}, Z < \frac{\mathcal{D}_{\hat{i}}\gamma_{\hat{i}}}{\Psi_{i}} \right].
 \end{align}
 Following \eqref{jjjsg_sat}, the CDF $F_{\gamma_{i,\hat{i}}^\text{sat}}({\gamma_{\hat{i}}})$ can be expressed in integration form as
 \begin{align}\label{kmklw_sat}
 F_{\gamma_{i,\hat{i}}^\text{sat}}(\gamma_{\hat{i}})&= \int^{\iota_4}_{z=\frac{\mathcal{D}_{\hat{i}}\gamma_{\hat{i}}}{\Psi_{i}}}\!\!\!f_{Z}(z)
 \!\!\int^{\frac{\mathcal{C}_{i} \gamma_{\hat{i}}z}{\Psi_{i} z-\mathcal{D}_{\hat{i}}\gamma_{\hat{i}} }}_{w=P_\text{th}} \!\!f_{W}\!(\!w\!)dw\,dz \nonumber\\
 &+ \int^{\frac{\mathcal{D}_{\hat{i}}\gamma_{\hat{i}}}{\Psi_{i}} }_{z=0}\!\!\!f_{Z}(z)
 \!\!\int^{\infty}_{w=P_\text{th}} \!\!f_{W}\!(\!w\!)dw\,dz.
 \end{align}
On substituting the respective PDFs and applying the required mathematical formulations \cite[eqs.  3.381.1, 3.471.9]{math}, \eqref{kmklw} and \eqref{kmklw_sat} can be expressed as in Lemma \ref{lem2}.

\section*{Acknowledgment}
The authors would like to thank the editor and anonymous reviewers for their valuable comments and suggestions, which helped to improve the clarity and quality of the paper.


\begin{thebibliography}{40}
	\bibliographystyle{IEEEtran}
	
	\bibitem{Saxena}
	M. Agiwal, A. Roy, and N. Saxena, \newblock{\textquotedblleft Next generation 5G wireless networks: A comprehensive survey,\textquotedblright} \emph{\newblock IEEE Commun. Surveys \& Tutorials}, vol. 18, no. 3, pp. 1617-1655, thirdquarter 2016.
	
	
	\bibitem{Khan2017}
	A. A. Khan, M. H. Rehmani, and A. Rachedi, \newblock{\textquotedblleft Cognitive-radio-based Internet of things: Applications, architectures, spectrum related functionalities, and future research directions,\textquotedblright} \emph{\newblock IEEE Wireless Commun.}, vol. 24, no. 3, pp. 17-25, June 2017.
	
	
	\bibitem{Goldsmith2009}
	A. Goldsmith, S. A. Jafar, I. Maric, and S. Srinivasa, \newblock{\textquotedblleft Beaking spectrum gridlock with cognitive radios: An information thoertic perspective,\textquotedblright} \emph{\newblock Proc. of the IEEE}, vol. 97, no. 5, pp. 894-914, May 2009.
	
	\bibitem{Sudevalayam}
	S. Sudevalayam and P. Kulkarni, \newblock{\textquotedblleft Energy harvesting sensor nodes: Survey and implications,\textquotedblright} \emph{\newblock IEEE Commun. Surveys and Tutorials}, vol. 13, no. 3, pp. 443-461, Sep. 2011.
	%
	
	
	\bibitem{Nasir}
	A. A. Nasir, X. Zhou, S. Durrani, and R. A. Kennedy, \newblock{\textquotedblleft Wireless-powered relays in cooperative communications: Time-switching relaying protocols and throughput analysis,\textquotedblright} \emph{\newblock IEEE Trans. Commun.}, vol. 63, pp. 1607-1622, May 2015.
	
	\bibitem{Varshney}
	L. R. Varshney, \newblock{\textquotedblleft Transporting information and energy simultaneously,\textquotedblright} \emph{\newblock Proc. IEEE ISIT}, pp. 1612-1616, July 2008.
	
	\bibitem{Zhou}
	X. Zhou, R. Zhang, and C. K. Ho, \newblock{\textquotedblleft Wireless information and power transfer: Architecture design and rate-energy tradeoff,\textquotedblright} \emph{\newblock IEEE Trans. Commun.}, vol. 61, no. 11, pp. 4754-4767, Nov. 2013.
	
	\bibitem{LWang2017}
	L. Wang, F. Hu, Z. Ling, and B. Wang, \newblock{\textquotedblleft Wireless information and power transfer to maximize information throughput in WBAN,\textquotedblright} \emph{\newblock IEEE Internet of Things J.}, vol. 4, no. 5, pp. 1663-1670, Oct. 2017.
	
		\bibitem{ydong}
	Y. Dong, M. J. Hossain, and J. Cheng, \newblock{\textquotedblleft Performance of wireless powered amplify and forward relaying over Nakagami-$m$ fading channels with nonlinear energy harvester, \textquotedblright} \emph{\newblock IEEE Commun. Lett.}, vol. 20, no. 4, pp. 672-675, April 2016.
	
	\bibitem{DSGVTC}
	D. S. Gurjar and H. H. Nguyen \newblock{\textquotedblleft Bidirectional primary and secondary transmissions with hybrid-SWIPT in  cognitive radio networks,\textquotedblright} \emph{\newblock IEEE VTC 2018-Fall}, Chicago, USA, August 27-30, 2018.
	
	\bibitem{TLi}
	T. Li, P. Fan, and K. B. Letaief, \newblock{\textquotedblleft Outage probability of energy harvesting
		relay-aided cooperative networks over rayleigh fading channel,\textquotedblright} \emph{\newblock IEEE Trans. Veh. Technol.}, vol. 65, pp. 972-978, Jan. 2016.
	
	\bibitem{HLee}
	H. Lee, C. Song, S.-H. Choi, and I. Lee, \newblock{\textquotedblleft Outage probability analysis
		and power splitter designs for SWIPT relaying systems with direct link,\textquotedblright} \emph{\newblock IEEE Commun. Lett.}, Nov. 2016.
	
	\bibitem{Men}
	J. Men, J. Ge, C. Zhang, and J. Li, \newblock{\textquotedblleft Joint optimal power allocation and relay selection scheme in energy harvesting asymmetric two-way relaying system,\textquotedblright} \emph{\newblock IET Commun.}, vol. 9, no. 11, pp. 1421-1426, July 2015.
	
	
	\bibitem{Du}
	G. Du, K. Xiong, Y. Zhang, and Z. Qiu, \newblock{\textquotedblleft Outage analysis and optimization for time switching-based two-way relaying with energy harvesting relay node,\textquotedblright} \emph{\newblock KSII Trans. Internet and Info. Systems}, vol. 9, no. 2, pp. 545-563, Feb. 2015.
	
	
	
	\bibitem{Hu}
	R. Hu and T.-M. Lok, \newblock{\textquotedblleft Power splitting and relay optimization for two-way relay SWIPT systems,\textquotedblright} \emph{\newblock Proc. IEEE ICC}, Malaysia, May 2016.
	
	
	\bibitem{Peng}
	C. Peng, F. Li, and H. Liu, \newblock{\textquotedblleft Optimal power splitting in two-way
		decode-and-forward relay networks,\textquotedblright} \emph{\newblock IEEE Commun. Lett.}, vol. 21, no. 9, pp. 2009-2012, Sep. 2017.
	
	\bibitem{Do}
	T. P. Do, I. Song, and Y. H. Kim, \newblock{\textquotedblleft Simultaneous wireless transfer of power and information in a decode-and-forward two-way relaying network,\textquotedblright} \emph{\newblock IEEE Trans. Wireless Commun.}, vol. 16, no. 3,  pp. 1579-1592, Mar. 2017.
	
	\bibitem{SourabhTVT}
    S. Solanki, V. Singh, and P. K. Upadhyay, \newblock{\textquotedblleft RF energy harvesting in hybrid two-way relaying systems with hardware impairments,\textquotedblright} \emph{\newblock IEEE Trans. Veh. Technol.}, vol. 68, no. 12, pp. 11792-11805, Dec. 2019.
	
	
	\bibitem{Yin2014}
	S. Yin, E. Zhang, Z. Qu, L. Yin, and S. Li, \newblock{\textquotedblleft Optimal cooperation strategy in cognitive radio systems with energy harvesting,\textquotedblright} \emph{\newblock IEEE Trans. Wireless Commun.}, vol. 13, no. 9, pp. 4693-4707, Sep. 2014.
	
	
	
	\bibitem{Wang2016}
	Z. Wang, Z. Chen, B. Xia, L. Luo, and J. Zhou, \newblock{\textquotedblleft Cognitive relay networks with energy harvesting and information transfer: Design, analysis, and optimization,\textquotedblright} \emph{\newblock IEEE Trans. Wireless Commun.}, vol. 15, no. 4, pp. 2562-2576, Apr. 2016.
	
	\bibitem{Im2015}
	G. Im and J. H. Lee, \newblock{\textquotedblleft Outage probability of underlay cognitive radio networks with SWIPT-enabled relay,\textquotedblright} \emph{\newblock Proc. IEEE VTC 2015-Fall},  Boston, MA, 2015, pp. 1-5.
	
	\bibitem{Yang2016}
	Z. Yang, Z. Ding, P. Fan, and G. K. Karagiannidis, \newblock{\textquotedblleft Outage performance of cognitive relay networks with wireless information and power transfer,\textquotedblright} \emph{\newblock IEEE Trans. Veh. Technol.}, vol. 65, no. 5, pp. 3828-3833, May 2016.
	
	\bibitem{Kalamkar2017}
	S. S. Kalamkar and A. Banerjee, \newblock{\textquotedblleft Interference-aided energy harvesting: Cognitive relaying with multiple primary transceivers,\textquotedblright} \emph{\newblock IEEE Trans. Cognit.	Commun. Netw.}, vol. 3, no. 3, pp. 313-327, Sept. 2017.
	
	\bibitem{Verma2017}
	D. K. Verma, R. Y. Chang, and F. T. Chien, \newblock{\textquotedblleft Energy-assisted decode-and-forward for energy harvesting cooperative cognitive networks,\textquotedblright} \emph{\newblock IEEE Trans. Cognit.
		Commun. Netw.}, vol. 3, no. 3, pp. 328-342, Sep. 2017.
	
	\bibitem{Yan2017}
	J. Yan and Y. Liu, \newblock{\textquotedblleft A dynamic SWIPT approach for cooperative cognitive radio networks,\textquotedblright} \emph{\newblock IEEE Trans. Veh. Technol.}, vol. 66, no. 12, pp. 11122-11136, Dec. 2017.
	
	\bibitem{Nguyen2018}
	B. V. Nguyen, H. Jung, D. Har, and K. Kim, \newblock{\textquotedblleft Performance analysis of a cognitive radio network with an energy harvesting secondary transmitter under Nakagami-${m}$ fading,\textquotedblright} \emph{\newblock IEEE Access}, vol. 6, pp. 4135-4144, 2018.
	
	
\bibitem{Mukherjee}
A. Mukherjee, T. Acharya, and M. R. A. Khandaker, \newblock{\textquotedblleft Outage analysis for SWIPT-enabled two-way cognitive cooperative communications,\textquotedblright} \emph{\newblock IEEE Trans. Veh. Technol.}, vol. 67, no. 9, pp. 9032-9036, Sept. 2018.
	
	
\bibitem{Zhou2018}
X. Zhou and Q. Li, \newblock{\textquotedblleft Energy efficiency for SWIPT in MIMO two-way amplify-and-forward relay networks,\textquotedblright} \emph{\newblock IEEE Trans. Veh. Technol.}, vol. 67, no. 6, pp. 4910-4924, June 2018.

\bibitem{Tang2018}
J. Tang, A. Shojaeifard, D. K. C. So, K. Wong, and N. Zhao, \newblock{\textquotedblleft Energy efficiency optimization for CoMP-SWIPT heterogeneous networks,\textquotedblright} \emph{\newblock IEEE Trans. Commun.}, vol. 66, no. 12, pp. 6368-6383, Dec. 2018.
	
\bibitem{DSIoT}
D. S. Gurjar, H. H. Nguyen, and H. D. Tuan, \newblock{\textquotedblleft Wireless information and power transfer for IoT applications in overlay cognitive radio networks,\textquotedblright} \emph{\newblock IEEE Internet of Things J}. vol. 6, no. 2, pp. 3257-3270, April 2019.

\bibitem{Ye2019}
J. Ye, Z. Liu, H. Zhao, G. Pan, Q. Ni, and M. Alouini, \newblock{\textquotedblleft Relay selections for cooperative underlay CR systems with energy harvesting,\textquotedblright} \emph{\newblock IEEE Trans. Cognitive Commun. \& Netw.},  vol. 5, no. 2, pp. 358-369, June 2019.

\bibitem{Kishore2019}
R. Kishore, S. Gurugopinath, P. C. Sofotasios, S. Muhaidat, and N. Al-Dhahir, \newblock{\textquotedblleft Opportunistic ambient backscatter communication in RF-powered cognitive radio networks,\textquotedblright} \emph{\newblock IEEE Trans. Cognitive Commun. \& Netw.},  vol. 5, no. 2, pp. 413-426, June 2019.


\bibitem{Deng2019}
Z. Deng, Q. Li, Q. Zhang, L. Yang, and J. Qin, \newblock{\textquotedblleft Beamforming design for physical layer security in a two-way cognitive radio IoT network with SWIPT,\textquotedblright} \emph{\newblock IEEE Internet of Things J.},  vol. 6, no. 6, pp. 10786-10798, Dec. 2019.

\bibitem{Singh2020}
A. Singh, M. R. Bhatnagar, and R. K. Mallik, \newblock{\textquotedblleft Secrecy outage performance of SWIPT cognitive radio network with imperfect CSI,\textquotedblright} \emph{\newblock IEEE Access},  vol. 8, pp. 3911-3919, 2020.

\bibitem{Zhang2020}
Z. Zhang, Y. Lu, Y. Huang, and P. Zhang, \newblock{\textquotedblleft Neural network-based relay selection in two-way SWIPT-enabled cognitive radio networks,\textquotedblright} \emph{\newblock IEEE Trans. Veh. Technol.},  vol. PP, no. 99, April 2020.



\bibitem{Shi2020}
L. Shi, W. Cheng, Y. Ye, H. Zhang, and R. Q. Hu, \newblock{\textquotedblleft Heterogeneous power-splitting based two-way DF relaying with non-linear energy harvesting,\textquotedblright} \emph{\newblock IEEE Global Communications Conference (GLOBECOM)},  Abu Dhabi, United Arab Emirates, 2018, pp. 1-7.


\bibitem{Liu2020}
Y. Liu, Y. Ye, H. Ding, F. Gao, and H. Yang, \newblock{\textquotedblleft Outage performance analysis for SWIPT based	 incremental cooperative NOMA networks with non-linear harvester,\textquotedblright} \emph{\newblock IEEE Commun. Lett.},  vol. 24, no. 2, pp. 287-291, Feb. 2020.








%
	
		

	

	%
	%
	%
	%
	%

	\bibitem{math}
	I. Gradshteyn and I. Ryzhik, \newblock{\textquotedblleft Table of integrals, series, and products,\textquotedblright} \emph{\newblock Academic Press, San Diego, California,} 7th ed., 2007.
	
	\bibitem{LiuIET}
	Y. Liu, L. Wang, M. Elkashlan, T. Q. Duong, and A. Nallanathan, \newblock{\textquotedblleft Two-way relay networks with wireless power transfer: Design and performance analysis,\textquotedblright} \emph{\newblock IET Commun.}, vol. 10, no. 14, pp. 1810-1819, June 2016.
	
    \bibitem{Shi}
    Y. Shi and R. Eberhart, \newblock{\textquotedblleft A modified particle swarm optimizer,\textquotedblright} \emph{\newblock IEEE International Conference on Evolutionary Computation Proceedings (Cat. No.98TH8360)}, pp. 69-73, Anchorage, AK, USA, 1998.
	

	
	
	
	\bibitem{cover1}
	M. T. Cover and J. A. Thomas, \emph{Elements of information theory}, Wiley-Interscience, New York, 1991.
	

	

	
	\bibitem{yli}
	Y. Li, H. Long, M. Peng and W. Wang, \newblock{\textquotedblleft Spectrum sharing with analog network coding,\textquotedblright} \emph{\newblock IEEE Trans. Veh. Technol.}, vol. 63, no. 4, pp. 1703-1716, May 2014.
	
	\bibitem{PK2017}
	P. K. Sharma and P. K. Upadhyay, \newblock{\textquotedblleft Performance analysis of cooperative spectrum sharing with multiuser two-way relaying over fading channels,\textquotedblright} \emph{\newblock IEEE Trans. Veh. Technol.}, vol. 66, no. 2, pp. 1324-1333, Feb. 2017.
	
    \bibitem{alsharoa2017optimization}
     A.~Alsharoa, H.~Ghazzai, A.~E. Kamal, and A.~Kadri, ``Optimization of a power
		splitting protocol for {two-way} multiple energy harvesting relay system,''
		\emph{IEEE Trans. Green Commun. and Networking}, vol.~1, no.~4, pp.	444--457, 2017.	
	
\bibitem{atta}
S. Atapattu and J. Evans, \newblock{\textquotedblleft Optimal energy harvesting protocols for wireless relay networks,\textquotedblright} \emph{\newblock IEEE Trans. Wireless Commun.}, vol. 15, no. 8, pp. 5789-5803, Aug. 2016.
	

\end{thebibliography}
\end{document}